\documentclass[preprint,12pt]{elsarticle}

\usepackage[top=1in, bottom=1.25in, left=1in, right=1in]{geometry}
\usepackage[font=small,justification=justified,format=plain]{caption}
\usepackage[utf8]{inputenc}
\usepackage[T1]{fontenc}
\usepackage{xcolor}
\usepackage{lineno}
\usepackage{epsfig}         
\usepackage{cuted}
\usepackage{amsmath}
\usepackage{amssymb}
\usepackage{float}
\usepackage{xcolor}         
\usepackage[normalem]{ulem} 
\usepackage{multirow}
\usepackage{multicol}
\usepackage{subcaption}
\usepackage{lscape}
\usepackage{bm}
\usepackage{verbatim}
\usepackage{graphicx}
\usepackage{psfrag}
\usepackage{hyperref} 
\usepackage[export]{adjustbox}
\usepackage{soul}
\usepackage{setspace}
\usepackage{todonotes}

\definecolor{mm}{rgb}{0, 0, 1} 

\journal{International Journal of Mechanical Sciences}

\begin{document}

\begin{frontmatter}
\title{Ultra-thin normal-shear-coupled metabarrier for low-frequency underwater sound insulation}

\author{V. F. Dal Poggetto\corref{label1}\fnref{label1}}
\ead{vinicius.fonseca-dal-poggetto@univ-lille.fr}

\author{M. Miniaci\corref{label1}\fnref{label1}}
\ead{marco.miniaci@univ-lille.fr; marco.miniaci@gmail.com}

\fntext[label1]{Univ. Lille, CNRS, Centrale Lille, Junia, Univ. Polytechnique Hauts-de-France, UMR 8520 - IEMN - Institut d’Electronique de Microélectronique et de Nanotechnologie, F-59000 Lille, France}

\date{\today}

\begin{abstract}
\it
Underwater noise pollution caused by anthropogenic activities, such as offshore wind farms, significantly affects marine life, hindering the intra- and inter-specific interactions of many aquatic species. A common strategy to mitigate these effects is to enclose the noise source within a physical barrier to achieve acceptable noise levels in the surrounding region. Underwater barriers typically achieve noise reduction through sound absorption based on locally resonant systems (e.g., foam and bubble elements) or sound reflecting systems (e.g., air bubble curtains). Although locally resonant-based solutions can yield significant noise attenuation levels, their performance is usually narrow-band. On the other hand, sound-reflecting systems require wider dimensions, presenting poor performance in the low-frequency range. Thus, significant low-frequency underwater noise attenuation remains an open issue, especially when considering thin structures that perform over a broad frequency range. In this work, we present the design of thin metamaterial-based acoustic barriers whose underwater noise attenuation is owed to tailored anisotropic material properties. A topology optimization approach is used to obtain a unit cell that maximizes the coupling between normal stresses and shear strains (and vice-versa). The resulting metabarrier presents a sub-wavelength thickness-to-wavelength ratio in the low-frequency range (circa 1/70 below 1 kHz) and also high sound transmission loss values at higher frequencies (almost 100 dB above 2 kHz). We also investigate the effects of increased hydrostatic pressure, presenting structural modifications that enable real-world applications. The results presented in this work indicate an efficient metamaterial-based solution for the mitigation of underwater noise, suggesting a fruitful direction for the exploitation of anisotropy in acoustic insulation applications.
\end{abstract}


\end{frontmatter}

\thispagestyle{empty}


\renewcommand{\figurename}{Fig.}

\section{Introduction}

The social focus on marine renewable energy has renewed the interest of researchers and industry in new designs for compact and hydrostatic pressure resistant devices that yield effective underwater noise reduction, especially at low frequencies.
In these frequency ranges, increasing human activity in marine environments, such as offshore wind farms, tidal stream turbines, wave energy converters, shipping and military operations, is causing greater disruption to marine life, affecting the behavior of aquatic species by, compromising their communication, orientation, feeding, parental care, and prey detection skills~\cite{duarte2021soundscape}.
This effect is particularly grave in the low frequency range,
where the primary frequencies of hearing and vocalization of the mammalian auditory system are located~\cite{national2003ocean}.
However, reducing noise at low frequencies with insulating devices much thinner than the wavelength of incident waves (sub-wavelength behavior) has long been and still is a major scientific and technological challenge.
This is particularly true in underwater acoustics, where associated wavelengths are significantly longer than in air and fluid-structure interaction plays a significant role~\cite{ross2013mechanics, dong2023underwater}.
Thus, in order to efficiently reduce anthropogenic pressure in the underwater environment, the design of novel resilient barriers with significant sound insulation properties in the low frequency range is critical.

Whenever directly addressing the noise source is not possible, an alternative solution to reduce noise is to place barriers between the source and the areas requiring protection.
The primary approaches for underwater acoustic insulation materials involve designing the impedance match (or mismatch) of these barriers with respect to the acoustic impedance of water, resulting in (i) absorptive or (ii) reflective behavior~\cite{dong2020review}.
To achieve this, common design strategies include fluid- or solid-filled cavities, backings, resonators, periodic structures, and metamaterials -- composite materials made of periodic arrangements of sub-wavelength structures -- which have been widely explored to date~\cite{qu2022underwater, gao2022acoustic, croenne2025review}.

When impedance matching is exploited, the use of barriers made of polymeric materials enhanced with viscoelastic micro- or macro-inclusions, as well as viscoelastic coatings embedded with periodic air pockets~\cite{kuhl1950sound, hladky1991analysis, dong2020review, fu2021review} is among the most well known and diffuse absorptive approaches.
In this case, insulation performance results from intrinsic sound absorption derived from relaxation of molecular chains in the polymer matrix material and frequency-dependent modifications of the barrier vibration characteristics (wave velocity, resonance frequencies, and mode shapes).
Relaxation of molecular chains provides rather good performance at high frequencies, while its ability to absorb low-frequency sound remains limited~\cite{kuhl1950sound}.
To improve performance, cavities of various shapes have been explored to facilitate the conversion of acoustic waves to shear waves, a phenomenon known to improve absorption due to high shear damping present in some materials~\cite{lin2022sound}.
Analytical and numerical studies indicate that the shape and deformation of the inclusions play a crucial role; however, the operational frequency bandwidth is constrained by the resonance frequencies of the inclusions~\cite{ivansson2008numerical}.
To address this limitation, researchers have proposed optimized geometries and material compositions for resonating inclusions, along with tailored properties of the host material, to improve attenuation efficiency and broaden the operational frequency range~\cite{fu2021review}.

Although relatively well-performing at mid and high frequencies, these approaches perform poorly below $1$ kHz. 
Ensuring impedance matching between resonating elements (such as Helmholtz-like resonators) and water has shown potential as an alternative, at least in proximity of the resonance frequency of the resonators.
Although extensively studied in the context of airborne acoustics~\cite{romero2020design}, their application in underwater environments has been less explored.
Duan et al.~\cite{duan2021tunable} numerically proved that an underwater quasi-Helmholtz resonator, including rubber coatings inside its resonant cavity, can lead to perfect absorption in the $100$ to $300$ Hz frequency range.
Duan et al.~\cite{duan2023deep} developed a theoretical model to predict the sound absorption performance of a sub-wavelength absorber made of a perforated face plate, a fluid-filled square honeycomb core with interior rubber coating, and a fixed back plate based on the sound absorption theory of the micro-perforated panel and electroacoustic analogy.
Multiple peaks of perfect absorption below 1 kHz have been numerically reported.
Furthermore, Qu et al.~\cite{qu2022underwater} proposed a structured tungsten-polyurethane composite, impedance-matched to water, that achieves a slow longitudinal sound speed through the optimal distribution of Fabry-P\`erot resonances over a broad frequency range ($4$ to $20$ kHz).

Other absorption-driven approaches include air or air/water-filled Helmholtz resonators, multilayered coated plates to generate slow waves~\cite{zhang2018subwavelength}, the introduction of hard and air-filled backings~\cite{zhao2014backing}, and hydrosound dampers (nets with air-filled elastic balloons)~\cite{elmer2014new}.
In offshore installations, such as pile driving for wind farms~\cite{elzinga2019manuscript}, air-filled Helmholtz resonators have achieved noise reductions of approximately $20$ dB at a central excitation frequency of $140$ Hz~\cite{peng2018modelling}.
Wochner et al.~\cite{wochner2016underwater} measured up to a peak loss of $37$ dB around $100$ Hz (measured using average one-third octave bands) in a Helmholtz resonator-based system with two fluids (water and air).
The main drawbacks of these approaches include their narrow-band nature and the difficulty in accounting for the elastic deformation of the resonators induced by hydrostatic pressure or by the oblique incidence of the acoustic wave, which can affect the resonant behavior of the finite system~\cite{su2023finite}.

The second approach, which relies on a strong impedance mismatch between the barrier and surrounding water, primarily works by reflecting acoustic waves to prevent their propagation through a given medium.
Insulation of water sound through impedance mismatch has the advantage of a more broadband effectiveness compared to using materials with band gaps induced by local resonances.
Common solutions of this type include (i) arrays of one fluid immersed in another with contrasting properties, such as air bubbles in water, and (ii) mechanical casings that enclose the noise source, such as those used around eolian piles during driving, with the goal of confining the energy of the sound waves to the space between the noise source and the barrier.
Lucke et al.~\cite{lucke2011use} reported a peak-to-peak mean level difference of $14$ dB inside and outside an air curtain installed to protect porpoises from the installation of wooden piles in Denmark.
W{\"u}rsig et al. reported~\cite{wursig2000development} reductions of $8$--$10$ dB in the $400$--$800$ Hz bands and $15$--$20$ dB in the $1.6$--$6.4$ kHz bands, indicating significantly poorer performance at lower frequencies.
Although commonly used, achieving attenuation in the very low frequency range ($< 600$ Hz) may require air curtains with bubbles ranging in radius from $8$ mm near the surface to $50$ mm at a depth of $30$ m~\cite{tsouvalas2020underwater}.
This poses a significant limitation for the practical application of this concept because of the challenges of generating and maintaining air bubbles in a controlled manner in an offshore environment.

Although shown in the ultrasonic regime, it is worth mentioning the works by Brunet et al., who observed ultra-slow Mie resonances in meta-fluids made of concentrated suspensions of macro-porous micro-beads engineered using soft matter techniques~\cite{brunet2015soft, tallon2021experimental}, and Leroy et al., who measured good acoustic attenuation in liquid foams over a broad high frequency range ($60$--$600$ kHz), proving through a theoretical model that the attenuation mechanism is derived from the existence of two non-dispersive bands in the dispersion relation for longitudinal acoustic waves, separated by a negative density regime~\cite{pierre2014resonant, pierre2017investigating}.

Architected acoustic panels represent another potential solution to reduce underwater noise acting as mechanical casings placed around noise sources.
However, water belongs to high-impedance media ($Z_w \approx 1.5 \times 10^6$ N $\cdot$ s/m$^3$), implying that barriers aimed at reflections due to a higher impedance with respect to water cannot efficiently isolate sound at low frequencies.
For example, for the case of a 50 mm thick steel plate immersed in water, more than $85$\% of the energy is transmitted if the frequency of the incident acoustic wave is below 500 Hz.

In the case of underwater casing systems for mitigation noise associated with pile installation, these are completely enclosed along the column of seawater.
For instance, Jansen et al.~\cite{jansen2012experimental} proposed noise mitigation screens made of double-walled steel cylinder filled with air submerged up to $25$ m of depth.
The proposed system works with combinations of (i) water or air/water between the pile and the casing and (ii) water or air inside the casing, yielding a frequency-dependent acoustic insertion loss (difference between transmission without and with the barrier) with a slope of about $2$ dB/octave, exceeding $20$ dB above $4$ kHz, being limited to $8$--$11$ dB when averaged over a large frequency range ($50$ Hz -- $40$ kHz).
However, the authors also state that the attenuation is less pronounced in the low frequency 
range ($100$--$250$ Hz).
Other technological solutions for underwater casing systems are also reported, although with a lesser degree of maturity in terms of practical deployment~\cite{koschinski2013development}.



Wang et al.~\cite{wang2022topological} introduced a topological optimization approach for lattice materials to create low acoustic impedance media, allowing "acoustic soft boundaries" compared to water.
By optimizing the topology of the material and applying a homogenization method, key parameters for effective low-frequency sound insulation have been identified.
The results revealed that minimal acoustic impedance depends not only on the degree of anisotropy but also on specific structural features that facilitate deformation.
Two designs with ultra-thin and deep sub-wavelength structures have been proposed relying on a bi-mode lattice material possessing (i) strong anisotropy and a positive Poisson's ratio or (ii) weak anisotropy and a negative Poisson's ratio.
Chen et al.~\cite{chen2020highly} explored the reflection of underwater sound waves on orthotropic solids, whose acoustic impedance can be customized through mechanical anisotropy, principal axes orientation, and velocity ratio of the quasi-transverse and quasi-longitudinal waves supported by these materials.
Numerical simulations have shown an optimal condition of an almost perfect reflection (97.7\%) of the incident acoustic energy in the case of normal incidence with an overall thickness of the lattice being two orders of magnitude smaller than the water wavelength in the $1.5$ -- $3.5$ kHz frequency range.
Wang et al.~\cite{wang2023acoustically} measured a loss of sound transmission of approximately $16$ dB within the $400$ -- $1200$ Hz frequency range, overcoming the mutual exclusion of low acoustic impedance and high mechanical properties by regulating the lattice orientation and incorporating a hierarchical morphology in an anisotropic metamaterial.
Wang et al.~\cite{wang2024topology} proposed reverse engineering of chiral metamaterials with optimized low acoustic impedance and appropriate stiffness using a topology optimization approach, demonstrating high and broadband sound transmission loss (STL) within the $1$ to $5$ kHz frequency range, as well as sufficient stiffness to ensure stable acoustic performance under moderate hydrostatic pressure.
Experiments carried out in a water-filled impedance tube revealed, on average, a reduction of more than $95\%$ of the incident sound energy.

Despite the approaches described above, achieving broadband and efficient underwater noise reduction at low frequencies (below 1000 Hz) remains a challenge, especially when using highly sub-wavelength structures that must also withstand hydrostatic pressure.
In this work, we address this challenge by introducing a periodic metamaterial with a unit cell designed to enable a tailored interaction between normal and shear behavior along predefined directions.
An elasticity-based approach allowed us to ensure efficient underwater acoustic waves reflection in broad frequency ranges.
We start by demonstrating the effect that a coupled normal--shear behavior has on fluid-immersed homogeneous anisotropic metabarriers.
Then, these results are utilized to formulate a topology optimization problem which maximizes such coupling.
Unlike existing approaches, our optimization problem is proposed in the static regime (zero frequency), depending only on the homogenized material properties of the structured material but not on the properties of the surrounding fluid.
The obtained results, however, are shown to be valid to non-zero frequencies under various situations.

The paper is organized as follows:
Section~\ref{sec_models_methods} presents the mathematical models and numerical methods used to investigate the STL curves of fluid-immersed anisotropic metabarriers and the formulation of the topology optimization problem that yields the optimized unit cell.
Section~\ref{sec_results} presents the results of the optimization problem and its corresponding STL curves for an increasing number of unit cells, including suitable modifications to account for hydrostatic pressure. We also perform two-dimensional simulations to assess the acoustic attenuation effects for a point source.
In Section~\ref{sec_conclusions} we present our conclusions and an outlook for future work.


\section{Mathematical modeling and numerical methods} \label{sec_models_methods}

\subsection{Normal-shear coupled behavior in anisotropic media}

\begin{figure}[h!]
 \centering
 \includegraphics[scale=1]{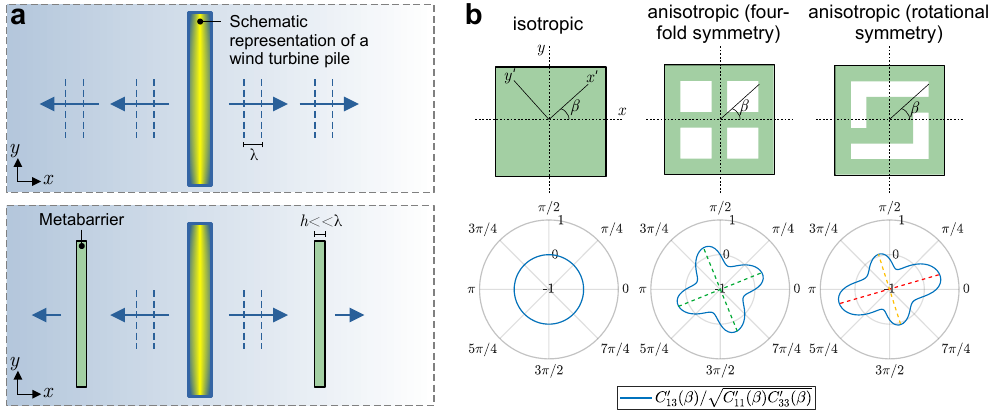}
 \caption{Underwater metabarrier.
 (\textbf{a}) A noise source produces acoustic waves (wavefronts represented in blue) with a typical wavelength $\lambda$ in an underwater medium (top panel).
 By enclosing the noise source within a metamaterial-based acoustic barrier (metabarrier, represented in green) with length $h \ll \lambda$, incident acoustic waves from the fluid partition at one side of the barrier are partially reflected, while a fraction of their energy is transmitted to the fluid partition at the other side (bottom panel). 
 (\textbf{b}) Typical examples of unit cells (top panel) with their respective behaviour (bottom panel) of isotropic materials (left panel), anisotropic materials with four-fold symmetry (center panel), and anisotropic materials with rotational symmetry (right panel), yielding direction-independent, $\pi/2$-rotational symmetric, and $\pi$-rotational symmetric constitutive matrix components. The green dashed line indicates the axis of symmetry (with respect to either reflection or rotation).
 }
 \label{figure1}
\end{figure}

Fig.~\ref{figure1}\textbf{a} reports a schematic representation of a noise source in an underwater medium generating acoustic waves with a typical wavelength $\lambda$ (top panel, Fig.~\ref{figure1}\textbf{a}).
Our objective is to design a metamaterial-based acoustic barrier (henceforth named "metabarrier") that interacts with the acoustic waves radiated by the noise source, reducing energy transmission at selected frequencies and achieving efficient underwater noise mitigation (bottom panel, Fig.~\ref{figure1}\textbf{a}).
Here, we do not account for energy absorption within the metabarrier.
To this end, we consider a metabarrier with an architected structure, whose periodicity is obtained by the repetition of a specially designed unit cell. The metabarrier presents a finite repetition of the unit cell in the $x$-direction (finite length $h \ll \lambda$) and an infinite number of unit cells in the $y$-direction (theoretically periodic medium).

To achieve the effect of significant STL values using an architected medium, we propose the exploitation of structural anisotropy.
To illustrate this behavior, let us consider a two-dimensional medium whose stress-strain relations are expressed in tensor notation as $\sigma_{ij} = C_{ijkl} \varepsilon_{kl}$, where $\sigma_{ij}$ is the Cauchy stress tensor, $C_{ijkl}$ is the fourth-order elasticity tensor and $\varepsilon_{kl}$ is the strain tensor~\cite{lai2009introduction}.
For a fixed Cartesian coordinate system with axes $xy$, these relations can be written in matrix form as
\begin{equation} \label{constitutive}
 \left\{
 \begin{array}{c}
  \sigma_x \\ \sigma_y \\ \tau_{xy}
 \end{array}
 \right\} = 
 \left[
 \begin{array}{ccc}
  C_{11} & C_{12} & C_{13} \\
  C_{21} & C_{22} & C_{23} \\
  C_{31} & C_{32} & C_{33} \\
 \end{array}
 \right]
\left\{
 \begin{array}{c}
  \varepsilon_x \\ \varepsilon_y \\ \gamma_{xy}
 \end{array}
 \right\} ,
\end{equation}
where $\sigma_x = \sigma_{xx}$ and $\sigma_y = \sigma_{yy}$ are the normal stresses, 
$\tau_{xy}$ is the shear stress, $C_{ij} = C_{ji}$, $\{i,j\} = \{1,2,3\}$, are terms of the constitutive matrix ($\mathbf{C}$),
$\varepsilon_x = \varepsilon_{xx}$ and $\varepsilon_y = \varepsilon_{yy}$ are normal strains, and $\gamma_{xy} = 2\epsilon_{xy}$ is the in-plane shear strain.
For a coordinate system $x^\prime y^\prime$ rotated with respect to $xy$ about an additional angle $\beta$ (Fig.~\ref{figure1}\textbf{b}, top left panel), it is possible to obtain the corresponding matrix in the new coordinate system using an adequate coordinate transformation matrix $\mathbf{T}$ of the form $\mathbf{C}^\prime = \mathbf{T}^T \mathbf{C} \mathbf{T}$ (see \ref{sec_constitutive} for details). As a consequence, each component of the direction-dependent elasticity tensor in matrix form $C_{ij}^\prime(\beta)$ can be analyzed separately.

The usual homogeneous structures employed in typical barriers~\cite{koschinski2013development} present isotropic constitutive behavior, yielding decoupled normal stresses and shear strains (and vice versa) for every direction. 
An example of a continuous homogeneous structure is shown in Fig.~\ref{figure1}\textbf{b} (top left panel), where we show (bottom left panel) the direction independence of the ratio between constitutive components $ C_{13}^\prime(\beta)/\sqrt{ C_{11}^\prime(\beta) C_{33}^\prime(\beta) }$.
In this case, the elasticity tensor components are invariable with respect to the orientation angle $\beta$ and $C_{13}^\prime(\beta) = C_{31}^\prime(\beta) = 0$, indicating that normal-shear coupling is not possible for any angle.

In the case of structures with an architected unit cell, however, an anisotropic behavior may be readily observed. Consider, for instance, the unit cell with four-fold symmetry shown in Fig.~\ref{figure1}\textbf{b} (top middle panel). 
In this case, the illustrated elasticity tensor components present a four-fold symmetry, displaying preferential angles ($\pi/8 + n \pi/4$, $n \in \mathcal{N}$, illustrated by the green dashed lines, bottom middle panel) for the coupling between normal and shear stresses ($C_{13}^\prime(\beta) \neq 0$). This implies, however, that no coupling is present at an arbitrary angle (e.g., $\beta=0$).

Finally, by removing the condition of four-fold symmetry, it is possible to design a unit cell that presents a non-zero value of normal-shear coupling at an arbitrary direction (e.g., $\beta=0$).
An example of a unit cell with such a condition is given in Fig.~\ref{figure1}\textbf{b} (top right panel), with its corresponding behavior showing $\pi/2$-rotational symmetry (dashed red and yellow lines, bottom right panel), and as a consequence, non-zero normal-shear coupling for the angle $\beta=0$.
This allows for the excitation of modes with a shear component due to the incidence of longitudinal (acoustic) modes from the fluid, thus hindering the efficiency of longitudinal wave mode propagation in the solid medium, reducing the transmission of energy between incident and transmitted acoustic waves.
Let us now verify the effect of a coupled normal-shear behavior on the STL curves of homogeneous anisotropic structures.

\subsection{Transfer matrix method} \label{sec_tmm}

\begin{figure}[h]
 \centering
 \includegraphics[scale=1]{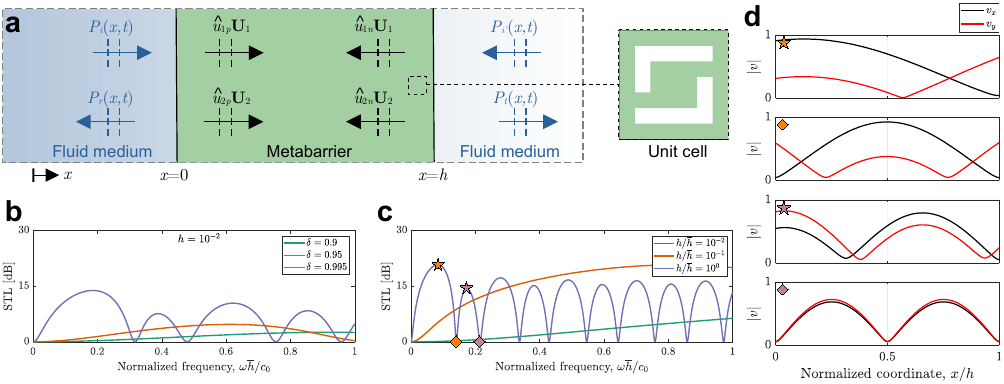}
 \caption{
 Investigation on the STL properties of a fluid-immersed homogeneous anisotropic metabarrier under normal incidence of acoustic waves.
 (\textbf{a}) Anisotropic homogeneous acoustic barrier with thickness $h$ subjected to incident waves ($P_i$, $P^\prime_i$), emitting reflected ($P_r$) and transmitted acoustic waves ($P_t$). The internal propagation of elastic waves is described by a pair of wave modes ($\mathbf{U}_1$, $\mathbf{U}_2$) with positive- and negative-going components (indices $p$ and $n$ for each amplitude $\hat{u}$, respectively).
 (\textbf{b}) Computed STL  in the normalized frequency range $\omega \overline{h} / c_0 \in [0,1]$ for the thickness $h=\overline{h}=10^{-2}$ and coupling factor $\delta=\{0.9, 0.95, 0.995\}$.
 (\textbf{c}) Same as (\textbf{a}) for $\delta=0.999$ and $h/\overline{h} = \{ 10^{-2}, 10^{-1}, 10^{0} \}$.
 The first and second peaks (dips) are represented by orange and purple stars (diamonds), respectively, for the case $h=10^0$.
 (\textbf{d}) Amplitude of normalized longitudinal ($v_x$) and transverse ($v_y$) velocities for the peaks and dips indicated in (\textbf{c}).
 }
 \label{figure2}
\end{figure}

To investigate the effects of coupled normal-shear behavior, we utilize a transfer matrix method~\cite{jimenez2021acoustic} to relate quantities of interest (in this case, acoustic pressures) on both edges of an acoustic barrier with homogeneous anisotropic material properties and a thickness $h$ in the $x$-direction (see Fig.~\ref{figure2}\textbf{a}) immersed in a fluid, yielding the corresponding STL curves.

For the derivation of the transfer matrix method, we consider waves propagating in the $x$-direction (same as the finite length of the metabarrier). 
The general form of structural displacements is written as
\begin{equation} \label{app_displacements_normal}
 \mathbf{u}(x,t) = e^{-\text{i} \omega t}
 ( \hat{u}_{1p} \mathbf{U}_1 e^{\text{i}k_1 x} +
   \hat{u}_{1n} \mathbf{U}_1 e^{-\text{i}k_1 x} +
   \hat{u}_{2p} \mathbf{U}_2 e^{\text{i}k_2 x} +
   \hat{u}_{2n} \mathbf{U}_2 e^{-\text{i}k_2 x} ) \, ,
\end{equation}
where $\mathbf{u} = \left\{ u_x, u_y \right\}^T$ is a displacement vector in Cartesian coordinates, $x$ is a coordinate whose origin is set at the fluid-structure interface between the metabarrier and incident acoustic waves,
$\text{i} = \sqrt{-1}$ is the imaginary unit number,
$\omega$ is the circular frequency, 
$t$ is the time coordinate,
$\mathbf{U}_1$ and $\mathbf{U}_2$ are generalized wave modes,
$k_1$ and $k_2$ their corresponding wavenumbers,
and $\hat{u}_{1p}$ and $\hat{u}_{1n}$ ($\hat{u}_{2p}$ and $\hat{u}_{2n}$) are, respectively, the complex amplitudes of waves traveling in the positive ($+x$) and negative ($-x$) directions, related to wave mode $\mathbf{U}_1$ ($\mathbf{U}_2$).
The values of complex amplitudes can be determined by setting appropriate boundary conditions at the fluid-structure interface.
Details on wavenumber and wave modes are given in~\ref{sec_model_wave_homogeneous}.

At the fluid-structure interfaces between metabarrier and surrounding fluid, shear stresses vanish~\cite{oudich2014general}, leading to a ratio of wave amplitudes given by
\begin{equation} \label{ratio_positive_negative}
 \left\{
    \begin{array}{c}
     \hat{u}_{1n} \\
     \hat{u}_{2n}
    \end{array}
 \right\} =
 \left[
    \begin{array}{cc}
     \Delta_{11} & \Delta_{12} \\
     \Delta_{21} & \Delta_{22}
    \end{array}
 \right]
 \left\{
    \begin{array}{c}
     \hat{u}_{1p} \\
     \hat{u}_{2p}
    \end{array}
 \right\} \, ,
\end{equation}
where
\begin{equation}
\begin{aligned}
 \Delta_{11} &= \frac{e^{-\text{i} k_2 h} - e^{\text{i} k_1 h}}{e^{-\text{i} k_2 h} - e^{-\text{i} k_1 h}} \, , 
 \Delta_{12} = \frac{C_z^{\phi_2} k_2}{C_z^{\phi_1} k_1} \frac{e^{-\text{i} k_2 h} - e^{\text{i} k_2 h}}{e^{-\text{i} k_2 h} - e^{-\text{i} k_1 h}} \, , \\
 \Delta_{21} &= \frac{C_z^{\phi_1} k_1}{C_z^{\phi_2} k_2} \frac{e^{-\text{i} k_1 h} - e^{\text{i} k_1 h}}{e^{-\text{i} k_2 h} - e^{-\text{i} k_1 h}} \, , 
 \Delta_{22} = \frac{e^{\text{i} k_2 h} - e^{-\text{i} k_1 h}}{e^{-\text{i} k_2 h} - e^{-\text{i} k_1 h}} \, ,
\end{aligned}
\end{equation}
with $C_z^{\phi} = C_{31} \cos \phi + C_{33} \sin \phi$, with $\phi_i = \angle \mathbf{U}_i$.
Details are given in~\ref{sec_derivation_tmm}.

The structural transfer matrix $\mathbf{T}^{(s)}$ allows to relate longitudinal displacements ($u_x$) and normal stresses ($\sigma_x$) at the right ($x=h$) and left ($x=0$) edges of the metabarrier as
\begin{equation} \label{transfer_matrix_s}
 \left\{
    \begin{array}{c}
     u_x \\
     \sigma_x
    \end{array}
 \right\}_{x=h}
 = \mathbf{T}^{(s)}
  \left\{
    \begin{array}{c}
     u_x \\
     \sigma_x
    \end{array}
 \right\}_{x=0} 
 = 
 \left[
    \begin{array}{cc}
     T^{(s)}_{11} & T^{(s)}_{12} \\
     T^{(s)}_{21} & T^{(s)}_{22}
    \end{array}
 \right]
 \left\{
    \begin{array}{c}
     u_x \\
     \sigma_x
    \end{array}
 \right\}_{x=0} ,
\end{equation}
where $T_{11}^{(s)} = T_{22}^{(s)}$, $\det(\mathbf{T}^{(s)}) = T_{11}^{(s)}T_{22}^{(s)} - T_{12}^{(s)}T_{21}^{(s)} = 1$ and $\mathbf{T}^{(s)}$ is real-valued. This matrix is computed as
\begin{equation}
    \mathbf{T}^{(s)} = \mathbf{M}(x=h) \mathbf{M}^{-1}(x=0) \, ,    
\end{equation}
where $\mathbf{M}(x)$ is obtained as
\begin{equation}
 \mathbf{M}(x) = 
 \left[
    \begin{array}{cccc}
     \cos \phi_1 e^{\text{i}k_1x} & \cos \phi_1 e^{-\text{i}k_1x} & \cos \phi_2 e^{\text{i}k_2x} & \cos \phi_2 e^{-\text{i}k_2x} \\
     C_x^{\phi_1} \text{i} k_1 e^{\text{i}k_1x} & -C_x^{\phi_1} \text{i} k_1 e^{-\text{i}k_1x} & C_x^{\phi_2} \text{i} k_2 e^{\text{i}k_2x} & -C_x^{\phi_2} \text{i} k_2 e^{-\text{i}k_2x} \\
    \end{array}
 \right]
 \left[
    \begin{array}{cc}
     1 & 0 \\
     \Delta_{11} & \Delta_{12} \\
     0 & 1 \\
     \Delta_{21} & \Delta_{22} \\
    \end{array}
 \right] \, ,
\end{equation}
with $C_x^{\phi} = C_{11} \cos \phi + C_{13} \sin \phi$.
Details are given in~\ref{sec_derivation_tmm}.

Let now the acoustic pressure waves (see Fig.~\ref{figure2}\textbf{a}) be denoted as
\begin{equation} \label{pressure_waves_x}
\begin{array}{cc}
 P_i(x,t) = e^{-\text{i} \omega t} \hat{P}_i e^{\text{i} k_0 x}, & 
 P_t(x,t) = e^{-\text{i} \omega t} \hat{P}_t e^{\text{i} k_0 (x-h) }, \\
 P_r(x,t) = e^{-\text{i} \omega t} \hat{P}_r e^{-\text{i} k_0 x}, &
 P_{i^\prime}(x,t) = e^{-\text{i} \omega t} \hat{P}_{i^\prime} e^{-\text{i} k_0 (x-h) },
\end{array}
\end{equation}
where
$P_i$ and $P_{i^\prime}$ are, respectively, incident waves propagating in the $+x$ and $-x$ directions, $P_r$ and $P_t$ are, respectively, reflected and transmitted waves (with respect to $P_i$), $k_0 = \omega / c_0$ is the acoustic wavenumber, and $c_0$ is the sound of speed in the fluid.
The pressure $P_{i^\prime}$ is introduced due to the necessity of symmetry in the formulation, and the shift $(x-h)$ in the exponential terms of the waves at the right end of the metabarrier is adopted for analytical convenience.

The continuity of accelerations and stresses at the fluid-structure interface between the metabarrier and surrounding fluid allows to write the acoustic transfer matrix $\mathbf{T}^{(a)}$, relating acoustic pressures on both sides of the metabarrier, as
\begin{equation}
 \left\{
    \begin{array}{c}
        \hat{P}_t \\
        \hat{P}_{i^\prime} \\
    \end{array}
 \right\}
 =
 \mathbf{T}^{(a)}
 \left\{
    \begin{array}{c}
        \hat{P}_i \\
        \hat{P}_r \\
    \end{array}
 \right\}
 =
 \left[
    \begin{array}{cc}
     T^{(a)}_{11} & T^{(a)}_{12} \\
     T^{(a)}_{21} & T^{(a)}_{22}
    \end{array}
 \right]
 \left\{
    \begin{array}{c}
        \hat{P}_i \\
        \hat{P}_r \\
    \end{array}
 \right\} \, ,
\end{equation}
where the terms of the acoustic transfer matrix $\mathbf{T}^{(a)}$ are given by
\begin{equation}
\begin{aligned}
 T^{(a)}_{11}  = T^{(s)}_{11} + \frac{\text{i}}{2} \bigg(-\frac{k_0}{\rho_0 \omega^2} T^{(s)}_{21} + \frac{\rho_0 \omega^2}{k_0} T^{(s)}_{12} \bigg), \, 
 &T^{(a)}_{12} = \frac{\text{i}}{2} \bigg(\frac{k_0}{\rho_0 \omega^2} T^{(s)}_{21} + \frac{\rho_0 \omega^2}{k_0} T^{(s)}_{12} \bigg), \\ 
 T^{(a)}_{21}  = - \frac{\text{i}}{2} \bigg(\frac{k_0}{\rho_0 \omega^2} T^{(s)}_{21} + \frac{\rho_0 \omega^2}{k_0} T^{(s)}_{12} \bigg), \, 
 &T^{(a)}_{22} = T^{(s)}_{22} + \frac{\text{i}}{2} \bigg(\frac{k_0}{\rho_0 \omega^2} T^{(s)}_{21} - \frac{\rho_0 \omega^2}{k_0} T^{(s)}_{12} \bigg) \, , \\ 
\end{aligned}
\end{equation}
with $\det(\mathbf{T}^{(a)}) = T_{11}^{(a)} T_{22}^{(a)} - T_{12}^{(a)}T_{21}^{(a)} = 1$ due to the reciprocity of the system.

Assuming no incident waves at the right edge of the metabarrier ($\hat{P}_{i^\prime}=0$) leads to a reflection coefficient $R = \hat{P}_r/\hat{P}_i = - T_{21}^{(a)} / T_{22}^{(a)}$, which allows to write the transmission coefficient as
\begin{equation}
 T = \frac{\hat{P}_t}{\hat{P}_i} = \frac{T_{11}^{(a)} T_{22}^{(a)} - T_{12}^{(a)} T_{21}^{(a)} }{T_{22}^{(a)}} = \frac{1}{T_{22}^{(a)}} \, .
\end{equation}

Finally, the STL can be computed as~\cite{kinsler2000fundamentals}
\begin{equation} \label{stl}
 \text{STL}(\omega) = 10 \log \bigg( \frac{\hat{P}_i}{\hat{P}_t} \bigg)^2 = 20 \log \bigg| \frac{1}{T} \bigg| \, .
\end{equation}

It is interesting to note that there exists a set of frequencies where the condition of full transmission (i.e., $|T|=1$, zero STL) is observed. This condition is achieved at Im$\{T\} = $Im$\Big\{ \frac{1}{T_{22}^{(a)}} \Big\}=0$, i.e.,
\begin{equation} \label{condition_T1}
    \frac{k_0}{\rho_0 \omega^2} T^{(s)}_{21} - \frac{\rho_0 \omega^2}{k_0} T^{(s)}_{12} = 0  \, ,
\end{equation}
which implies a real-valued $T_{22}^{(a)}$ at these frequencies (since $T_{22}^{(s)}$ is real-valued). As a consequence, at $|T|=1$, we have the equality $T_{22}^{(a)} = T_{22}^{(s)} = \pm 1$. As $\det(\mathbf{T}^{(s)}) = 1$ and $T_{22}^{(s)} = T_{11}^{(s)}$, this also implies in $T_{12}^{(s)} T_{21}^{(s)} = (T_{22}^{(s)})^2 - 1$. Since at these frequencies $T_{22}^{(s)} = \pm 1$, this also implies $T_{12}^{(s)} T_{21}^{(s)} = 0$, which combined with Eq.~(\ref{condition_T1}), also implies that both $T_{12}^{(s)} = 0$ and $T_{21}^{(s)} = 0$.
As a consequence, at each frequency where $|T|=1$ (zero STL), we have $\mathbf{T}^{(s)} = \mathbf{T}^{(a)} = \pm \mathbf{I}$, which implies $\hat{P}_t = \pm \hat{P}_i$, $u_x(x=h) = \pm u_x(x=0)$, and $\sigma_x(x=h) = \pm \sigma_x(x=0)$.
Thus, we conclude that full transmission conditions are associated with a formation of symmetric modes at the metabarrier.
A similar derivation of STL maxima condition is however less direct, since in these cases the condition $|T|=0$ is not achieved (which would imply infinite STL). We now proceed to utilize the derived transfer matrix method to evaluate the effect of couple normal-shear behavior on the STL curves.

\subsection{Effect of normal-shear coupling on STL curves} \label{sec_coupling_on_stl}

The proposed transfer matrix method can be used to present an illustrative example, computing the STL curves of a homogeneous two-dimensional structure whose constitutive matrix $\mathbf{C}$ presents components with values $C_{11} = 1$ GPa, $C_{33}/C_{11} = 0.9$ and a specific mass density of $\rho = 1000$ kg/m$^3$.
The ratio between the acoustic wave speed in the fluid and the solid is assumed as $c_0 / \sqrt{C_{11}/\rho} = 0.9$ and the ratio between their specific mass densities as $\rho_0 / \rho = 0.9$, thus similar to the ratio between the material properties of water and typical polymers.
We compute and investigate the STL curves for a barrier with a finite thickness $h$ over the normalized frequency range $\omega^* = \omega \overline{h} / c_0 \in [0,1]$, for $\overline{h}=10^{-2}$.

Following our previous discussion on anisotropy, we also define a mode coupling factor $\delta$ which relates normal stresses and shear strains (and vice-versa) as the the ratio between the $C_{13} = C_{31}$ terms of the constitutive matrix and the product $C_{11} C_{33}$ as
\begin{equation}
 \delta = \frac{ C_{13}  }{ \sqrt{ C_{11} C_{33} } } \, , 
\end{equation}
yielding $\delta \in (-1,1)$. In the case of an isotropic medium, $C_{13} = C_{31} = 0$ and one obtains $\delta=0$. 
This parameter is also typically associated to the degree of structural instability and used in the context of longitudinal-shear mode conversion~\cite{kweun2017transmodal}.


The STL curves computed for $h=10^0$ and $\delta = \{0.9, 0.95, 0.995 \}$ are shown in Fig.~\ref{figure2}\textbf{b}.
For the value $\delta = 0.9$, the presence of a local maximum of $2.6$ is noticed at $\omega^* = 0.83$.
For $\delta = 0.95$, we notice the presence of a maximum STL of $4.7$ located at $\omega^* = 0.55$, with the occurrence of a zero STL value (dip) at $\omega^* = 0.96$, which shows that increasing the value of $\delta$ presents the effects of (i) increasing the first peak STL value and (ii) red-shifting its frequency.
Finally, for $\delta = 0.995$, a total of $5$ STL peaks is noticed, also with an increase in the number of dips (total of $4$) in the same frequency range. Noticeably, the first peak is now located at $\omega^* = 0.17$ with an STL value of $13.8$. Thus, increasing $\delta$ from $0.9$ to $0.995$ leads to a (i) five-fold increase in the first STL peak and (ii) $80 \%$ reduction in its corresponding normalized frequency. As a consequence, an infinite STL value would be achieved close to the zero frequency, for $\delta \to 1^-$.

We also investigate the variation of the STL curves computed for the fixed value of $\delta = 0.999$ and increasing values of $h \in [10^{-2},10^0]$, as shown in Fig.~\ref{figure2}\textbf{c} (bottom panel)
For $h = 10^{-2}$, a maximum STL value of $7.0$ is computed at $\omega^*=1$, which suggests that even ultra-thin structures may present significant STL values, even if the first peak is located at higher frequencies. It is important to notice that remaining in the sub-wavelength regime (thickness much smaller than typical wavelength) is of interest, which then requires small values of $h$. On the other hand, sufficiently large values of $h$ are required to ensure that the homogenization hypotheses are valid, which may pose conflicting design objectives (sub-wavelength regime vs. homogenized material properties).
For $h = 10^{-1}$, a maximum STL value of $20.7$ is achieved at $\omega^*=0.75$, indicating the red-shift of STL peaks for an increase in the thickness.
Finally, for $h = 10^0$, a total of $10$ STL peaks and dips are observed in this frequency range, with a first STL peak of $20.7$ dB at $\omega^* = 0.075$. Interestingly, this represents a maintenance in the STL level of the first peak with respect to the case $h = 10^{-1}$, however with a ten-fold reduction in its frequency for a ten-fold thickness increase, thus suggesting that the first peak of the STL curve can be tuned using the thickness of the metabarrier.
Finally, we indicate the first two peaks (dips) in the STL curve for $h = 10^0$ using orange and purple stars (diamonds), respectively. The profiles of the absolute normalized velocities (and equivalently, displacements and accelerations) are shown in Fig.~\ref{figure2}\textbf{d}.

The modes corresponding to the selected peaks (orange and purple stars) present a significant decrease between the longitudinal velocities ($v_x$, black lines) at the left ($x=0$) and right ($x/h = 1$) ends of the metabarrier. This implies a decrease in the velocity (and acceleration) transmissibility  between the input and transmitted ends, which justifies the sharp peaks in the STL curves.
On the other hand, the modes corresponding to the dips in the STL curves (orange and purple diamonds) present symmetric profiles of longitudinal and transverse absolute velocities, resulting in unit transmissibility between the left and right edges of the metabarrier.
The $n$-th dip presents a phase variation of $n \pi$, $n \in \mathcal{N}$, resulting in complete in-phase or out-of-phase behavior between both edges of the metabarrier.
As a consequence, the transmitted and incident acoustic waves present the same amplitude, differing at most from a phase shift performed by the metabarrier.

\subsection{Optimization problem and unit cell design}

The previous discussion suggests that a suitable optimization objective to obtain the periodic unit cell that forms a metabarrier which yields maximum STL at a minimal operating frequency is the maximization of the mode coupling factor $\delta$. 
The desired value of $\delta \to 1^-$ would yield an infinite STL at zero frequency. However, this value cannot be achieved, as it represents a structural singularity. 

Thus, in order to obtain the periodic metamaterial unit cell, we propose a topology optimization problem which aims to minimize the objective function $\varphi$, given by
\begin{equation} \label{optimization_problem}
\begin{array}{rl}
 \underset{\mathbf{q} \in \{0,1\}^{n_q} }{\text{minimize}} & 
 \varphi(\mathbf{q}) = \big( |C_{13}| / \sqrt{C_{11}C_{33}} - 1 \big)^2 \, , \\
 \text{subjected to} & \omega_1 > \omega_{\min} \, ,
\end{array}
\end{equation}
where $\mathbf{q}$ is a binary vector composed of $n_q$ design variables representing the presence of material or voids in the unit cell, and
$\omega_1$ is the first resonant frequency of the system for free boundary conditions excluding rigid body motion, which must be above a certain threshold ($\omega_{\min}$).
The proposed objective function aims to minimize the distance between $|\delta|$ and $1$, while the first resonance frequency restriction is imposed to ensure a connectivity between the pixels of the candidate unit cell, thus discarding solutions that present disconnected (i.e., "floating") regions.

Each square unit cell used in the topology optimization process is described by a distribution of $2q \times 2q$ pixels.
The continuity between the pixels of contiguous unit cells is ensured by the correspondence of pixels between opposing edges.
Additionally, a $\pi/2$-rotational symmetry can be enforced in the internal region of the unit cell to ensure normal-shear coupling.
This yields a total of $n_q = 2q^2 - 2q -1$ pixels, which can then be encoded into a corresponding binary vector $\mathbf{q}$ (Fig.~\ref{figure3}\textbf{a}, middle panel). 
The represented unit cell can then be discretized using regular square finite elements under the assumption of plane strain~\cite{cook2007concepts}, allowing to compute the effective constitutive matrix of the homogenized unit using the method described in~\cite{pinho2009asymptotic}.
The dynamic behavior of the homogenized unit cell is valid in the long wavelength limit ($k \to \infty$, i.e., its validity decreases for increasing frequency values).
It is also important to note that since the homogenization procedure only concerns the static properties of the unit cell, the evaluation of binary vectors in the optimization problem (see Eq.~(\ref{optimization_problem})) presents a low computational cost, which allows to make an efficient use of genetic algorithms~\cite{goldberg2013genetic} to determine an optimal solution.

\section{Results} \label{sec_results}

\subsection{Result of optimization problem}

\begin{figure}[h]
 \centering
 \includegraphics[scale=1]{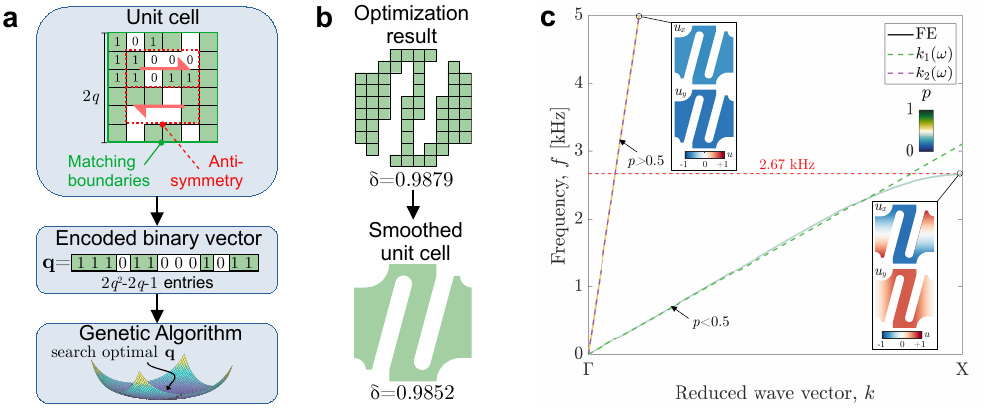}
 \caption{Optimization of the metabarrier unit cell.
 (\textbf{a}) Periodic unit cell (top panel) with a discretized geometry indicating $2q \times 2q$ pixels with either material ($1$) or void ($0$), with enforce $\pi/2$-rotational symmetry in the internal region (Anti-symmetry) and periodic boundaries (Matching boundaries). This description in encoded into a binary vector $\mathbf{q}$ (middle panel), which then is utilized in a genetic algorithm optimization (bottom panel).
 (\textbf{b}) Result of the optimization procedure (top panel), obtained for $q=5$, yielding a normalized stiffness ratio $\delta = 0.9879$. After obtaining a smooth approximate representation (bottom panel), this value is decreased to $\delta = 0.9852$.
 (\textbf{c}) Dispersion diagram (continuous lines) for the obtained unit cell (length $a=10$ mm), with polarization values $p$ indicating either completely longitudinal ($p=0$) or transverse motion ($p=1$). The wave modes are computed and shown at the X-point ($k = \pi/a$) of the first branch ($2.67$ kHz) and $k \approx 0.14 \pi/a$ point of the second branch ($5.00$ kHz). The dashed lines indicate the dispersion curves corresponding to the medium with effective homogenized properties for the first ($k_1$) and second branches ($k_2$).
 }
 \label{figure3}
\end{figure}

We apply the optimization procedure considering a structure composed of a plastic material with Young's modulus $E = 3$ GPa, Poisson's ratio $\nu = 0.3$, and specific mass density $\rho = 1150$ kg/m$^3$. The unit cell has a length  $a = 10$ mm, which is amenable to fabrication.
An example of a discretized unit cell, with a distribution of $2q \times 2q$ pixels, is shown in Fig.~\ref{figure3}\textbf{a}. The continuity between the pixels of contiguous unit cells and the internal $\pi/2$-rotational symmetry are represented as "Matching boundaries" (green contour) and "Anti-symmetry" (region inside the dashed red contour).

The result obtained using $q=5$, yielding a total of $2q \times 2q = 100$ pixels and $n_q = 39$ optimization variables (i.e., $2^{39} = 5.5 \times 10^{11}$ possible designs), and $\omega_{\min} = 0.1$ rad/s is presented in Fig.~\ref{figure3}\textbf{b} (top panel), displaying the discretized geometry of the unit cell and the obtained $\delta$ value ($\delta = 0.9879$). The obtained geometry then undergoes a smoothing procedure to yield a geometry amenable to fabrication (Fig.~\ref{figure3}\textbf{b}, bottom panel). As a result of the geometrical modifications, a slightly smaller value of $\delta$ is obtained ($\delta = 0.9852$). 

We then apply periodic Bloch-Floquet conditions to the resulting unit cell~\cite{mace2008modelling} and compute its dispersion diagram in the $\Gamma$X direction, with $\Gamma$ located at $k=0$ and X located at $k=\pi/a$.
For each wavenumber, we also compute a polarization metric given by
$p(u_x,u_y) = \int_S |u_y|^2 \, dS / \int_S |u_x|^2 + |u_y|^2 \, dS$,
where $u_x$ and $u_y$ are the $x$- and $y$-direction displacement components of a given wave mode and $S$ is the unit cell two-dimensional domain. As a result, $p=0$ ($p=1$) represents purely longitudinal (transverse) motion.
The band diagram computed using the FE method is shown in Fig.~\ref{figure3}\textbf{c} in the $[0,5]$ kHz frequency range using continuous lines, with polarization values indicated in a color scale ranging from $p=0$ (dark blue) to $p=1$ (dark green). 

The first obtained branch presents $p \in [0.403, 0.414]$, demonstrating a predominant longitudinal behavior ($p<0.5$), although with a high degree of hybridization between longitudinal and transverse motion. The wave  mode corresponding to this branch at the point X is also shown (colorbar represents displacement values, $u_x$ and $u_y$), illustrating this coupled behavior at $2.67$ kHz. Above this frequency (indicated with a horizontal red dashed line), the first branch is no longer present in the considered frequency range.
Likewise, the second branch presents $p \in [0.585, 0.589]$, exhibiting a predominant transverse behavior ($p>0.5$) with a significant hybridization between longitudinal and transverse motion. A wave mode corresponding to this branch is shown at $5$ kHz, displaying a rigid-body-like vertical motion due to the small wavenumber at this point ($k \approx 0.14 \pi/a$).
Therefore, it is possible to observe that even though the coupling between normal/shear stresses/strains was optimized in the static regime (i.e., $\omega = 0$), this also implies in coupled longitudinal-transverse motion . 

It is also interesting to note that the behavior predicted by the homogenized structure with effective properties (see~\ref{sec_model_wave_homogeneous}), represented by the dashed green and purple lines in Fig.~\ref{figure3}\textbf{c}, correlates well with the finite element-based solution up to $2.3$ kHz, when the first branch starts presenting a significant dispersive behavior. We now proceed to evaluate the STL properties of a metabarrier composed of repetitions of the unit cell obtained in the optimization procedure.

\clearpage
\subsection{Metabarrier STL curves}

\begin{figure}[h!]
 \centering
 \includegraphics[scale=1]{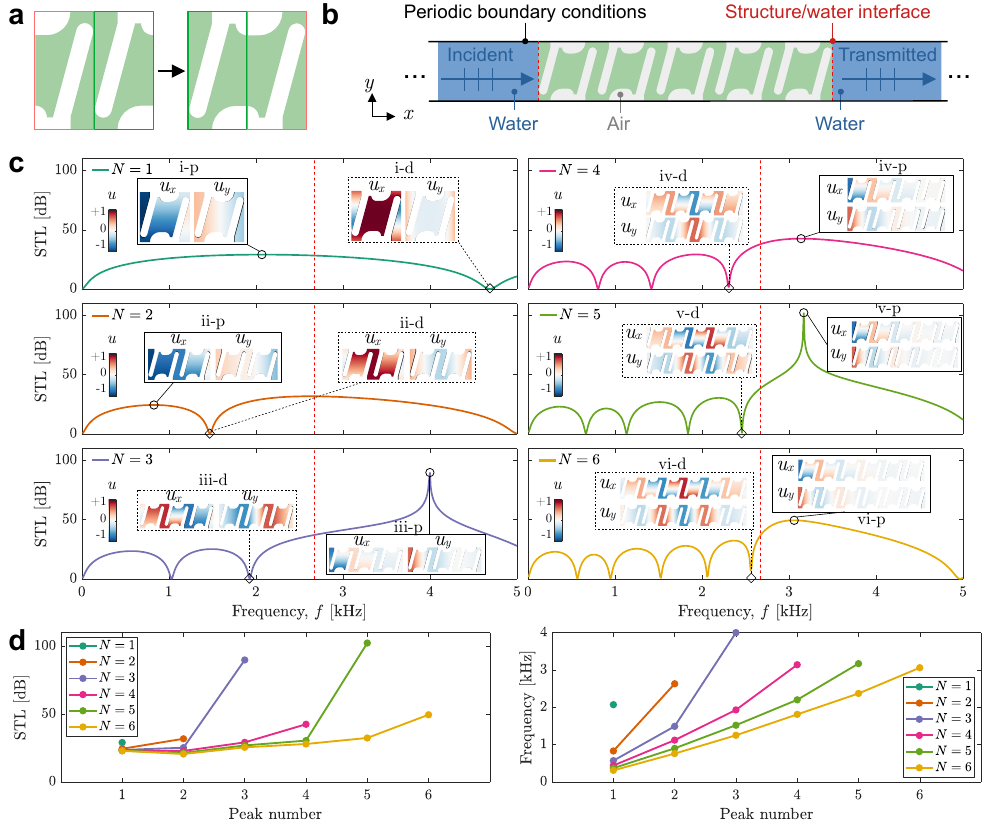}
 \caption{Metabarrier STL curves.
 (\textbf{a}) Redefinition of the unit cell (swapping regions inside the red and green contours, left to right) to obtain edges without gaps.
 (\textbf{b}) Sequential arrangement of unit cells filled with air and submerged into water. The medium is theoretically infinite in the $y$-direction, which allows to enforce periodic boundary conditions at the bottom and top edges of the overall unit cell. The ratio between the power of the transmitted and incident acoustic waves is then computed to obtain the STL curves.
 (\textbf{c}) STL curves for an increasing number of unit cells ($N=1,2,\cdots,6$).
 The peak (dips) displacement profiles are marked from i-p to vi-p (i-d to vi-d), shown in a normalized scale with respect to the largest absolute displacement.
 (\textbf{d}) Summary of values for STL peaks (left panel) and frequencies (right panel).
 }
 \label{figure4}
\end{figure}

The unit cell resulting from the optimization process (Fig.~\ref{figure3}\textbf{b}) presents inconvenient gaps in its left and right ends, resulting in finite structures with cavities at their edges.
To remedy this issue, we may swap the left and right halves of the unit cell keeping their orientation (regions inside the red and green contours in Fig.~\ref{figure4}\textbf{a}, respectively), which does not affect the dispersion relation of the material due to its periodicity. 
This unit cell is then repeated along the $x$-direction to obtain finite structures with a thickness equal to $h = Na$, where $N \in \mathcal{N}^*$ is the number of utilized unit cells.

The resulting finite structures are utilized to evaluate the STL (see Eq.~(\ref{stl})) when immersed in water, as illustrated in Fig.~\ref{figure4}\textbf{b}, considering a periodic medium in the $y$-direction.
The numerical procedure for computing the STL curves is described in~\ref{sec_stl_method_fe}, which we perform considering a solid structure filled with air fluid. The effect of the consideration of the internal fluid is a slight blue-shift in the STL curves with respect to the case without fluid (this comparison is presented in~\ref{sec_additional_results}). The material properties of water (air) are the specific mass density $\rho_w = 998$ kg/m$^3$ ($\rho_a = 1.2$ kg/m$^3$), with a speed of sound $c_w = 1481$ m/s ($c_a = 343$ m/s).
These results are shown in Fig.~\ref{figure4}\textbf{c} for an increasing number of unit cells, $N = 1,2,\cdots,6$.
Displacement profiles corresponding to selected peaks and dips are also shown considering longitudinal and transverse displacements ($u_x$ and $u_y$, respectively), normalized with respect to the largest computed displacement (i.e., $|u| = \sqrt{|u_x|^2+|u_y|^2}$).
The vertical dashed red line indicates the start of the frequency region where a single wave mode is present (corresponding to the horizontal dashed red line in Fig.~\ref{figure3}\textbf{c}).

We begin our analysis with $N=1$, a particularly interesting case due to its very thin profile. In the shown frequency range, this structure presents a maximum STL of $29.0$ dB at $2.07$ kHz (displacement profile i-p).
At this frequency, the structure presents a remarkable subwavelength performance, since the ratio between the thickness $h=a$ and the associated wavelength $\lambda = c_w/f$ yields the ratio $h/\lambda \approx 1/70$. The STL curve reaches a first dip at $4.7$ kHz, presenting a symmetric displacement profile with respect to the $y$-axis (displacement i-d), which justifies a unit ratio between displacements and also acoustic pressures, yielding a zero STL (a complete explanation is given in Section~\ref{sec_coupling_on_stl}).
However, no visible correlation can be described with respect to the dispersion diagram shown in Fig.~\ref{figure3}\textbf{c} due to the lack of periodicity in this case, which considers a single unit cell.

For the case $N=2$, a first STL peak of $24.4$ dB is computed at $0.83$ kHz, with a displacement profile (ii-p) indicating a behavior equivalent to that of the first peak computed at $N=1$ (i-p), however with a longer wavelength for both $u_x$ and $u_y$ (i.e., the displacement profiles become distributed along two unit cells, instead of one).
The second STL peak presents an attenuation of $31.8$ dB at $2.63$ kHz. We do not show this displacement profile since the formation of modes around $2.67$ kHz will be discussed ahead.
Also, a single dip is formed between these peaks at $1.47$ kHz, displaying a symmetric displacement profile with respect to the $y$-axis (ii-d), which has the same type of symmetry displayed by the first STL dip with $N=1$ (i-d).
Thus, it is possible to notice that similar behaviors are observed for the first peak and first dip for an increasing number of unit cells ($N= 1, 2 $), indicating the preservation of the behavior of equivalent peaks and dips, however with a significant red-shift in frequency.

For the case $N=3$, three STL peaks are computed with a $23.5$, $25.2$, and $89.8$ dB attenuation, at the frequencies  $0.57$, $1.49$, and $4.00$ kHz, respectively.
The first two peaks (not shown here for the sake of brevity) present a similar behavior to the previously computed peaks for the cases $N=1$ and $N=2$, however with a longer wavelength ($3$ unit cells instead of $1$ or $2$).
Also, the most notable difference between the first and the second peaks is the normalized longitudinal displacements, which change from half a wavelength, at the first peak, to three wavelengths, in the second peak.
Interestingly, the third peak, presenting a much more considerable attenuation, displays a displacement profile similar to a Bragg-scattering (iii-p), which can be explained due to the frequency range where it occurs (i.e., above $2.67$ kHz), corresponding to a band gap which is partial with respect to the first propagative branch (see Fig.~\ref{figure3}c). 
In this case, the dips are located at $1.03$ and $1.93$ kHz. It is interesting to notice that the displacement profile associated with the second dip (iii-d) displays a similar behavior to the wave mode computed at the X-point (see Fig.~\ref{figure3}c) at its center unit cell.

For the remaining considered cases ($N= 4, 5, 6$) we only report the dips that are closest to the frequency corresponding to the wave mode at the X-point ($2.67$ kHz) and the first peak observed immediately above this frequency. The corresponding dips (iv-d, v-d, and vi-d, respectively) present similar displacement profiles, with a behavior explained by the wave mode at the X-point of the first propagative branch (see Fig.~\ref{figure3}\textbf{c}).  
Indeed, for a sufficient number of unit cells, a vibration mode is expected precisely at the frequency corresponding to this wave mode, in the case of unit cells with vacuum in their cavities (situation corresponding to the computed band diagrams). We have verified this by computing the STL curve for a total of $N=10$ unit cells, checking the existence of the last dip at this region at $2.6$ kHz. 
It is also interesting to note that the displacement profiles corresponding to the peaks (iv-p, v-p, and vi-p, respectively) present similar behaviors to that of the dips, however with a spatial decay, which suggests a Bragg-scattering behavior considering the first branch of the dispersion diagram. The existence of the second branch at the same frequency range, however, impedes the formation of a complete band gap at this direction. 
We also note here the STL values and frequencies for the sequential peaks occurring at each case, respectively, 
(i) $N=4$: $23.2$, $22.8$, $29.1$, and $42.4$ dB at $0.44$, $1.12$, $1.93$, and $3.14$ kHz; 
(ii) $N=5$: $23.0$, $21.4$, $26.9$, $30.4$, and $102.3$ dB at $0.37$, $0.90$, $1.52$, $2.20$, and $3.17$ kHz; 
and (iii) $N=6$: $22.9$, $20.5$, $25.5$, $27.9$, $32.4$, and $49.5$ dB at $0.31$, $0.76$, $1.25$, $1.81$, $2.37$, and $3.06$ kHz.
At each case, the dips are formed at 
(i) $N=4$: $0.81$, $1.42$, and $2.31$ kHz; 
(ii) $N=5$: $0.67$, $1.13$, $1.84$, and $2.46$ kHz; and
(iii) $N=6$: $0.57$, $0.95$, $1.53$, $2.06$, and $2.56$ kHz.

We summarize the data concerning the attenuation and frequency of each maxima for the computed STL curves, respectively, in Figs.~\ref{figure4}\textbf{d} and \textbf{e}, for an increasing number of unit cells.
It is interesting to notice that, although for certain cases some peaks present a pronounced attenuation behavior similar to an anti-resonance (e.g., third peak for $N=3$ and fifth peak for $N=5$), the general trend concerning attenuation levels is to present a practically constant value. On the other hand, for a sufficient number of unit cells, the frequency at which these peaks occur presents an almost linear behavior starting from the first peak.

At this point, the question may arise whether the STL properties in this type of medium are only due to the high degree of hybridization observed ($p \approx 0.5$) in the wave modes shown by the dispersion diagram (Fig.~\ref{figure3}\textbf{c}).
To verify that this is not the case, we have performed an additional optimization procedure resulting in a structure with a perfect coupling between longitudinal and shear motion ($p \approx 0.5$ for all wave vectors in the $[0,5]$ khz frequency range).
In fact, the STL curves computed for this type of structure (see~\ref{app_fully_coupled}) present smaller peak values. Thus, this demonstrates that the fundamental aspect that guarantees large STL values is the mode coupling factor $\delta$.

\clearpage
\subsection{Influence of hydrostatic pressure on the design of the metabarrier}

After having presented the underlying physical phenomena responsible that generate significant STL in the low-frequency range, we now turn to a practical implementation concerning a fixed number of unit cells.
To this end, we proceed with the design of our metabarrier by choosing a fixed number of unit cells with considerable sub-wavelength thickness as well as remarkable STL features. 
We choose $N=3$ (see Fig.~\ref{figure4}\textbf{c}, middle left panel), whose thickness-to-wavelength ratio is $h/\lambda \approx 1/50$ in the operating frequency of $1$ kHz, also presenting a considerable STL ($\approx 90$ dB) at $4$ kHz.

\begin{figure}[h!]
 \centering
 \includegraphics[scale=1]{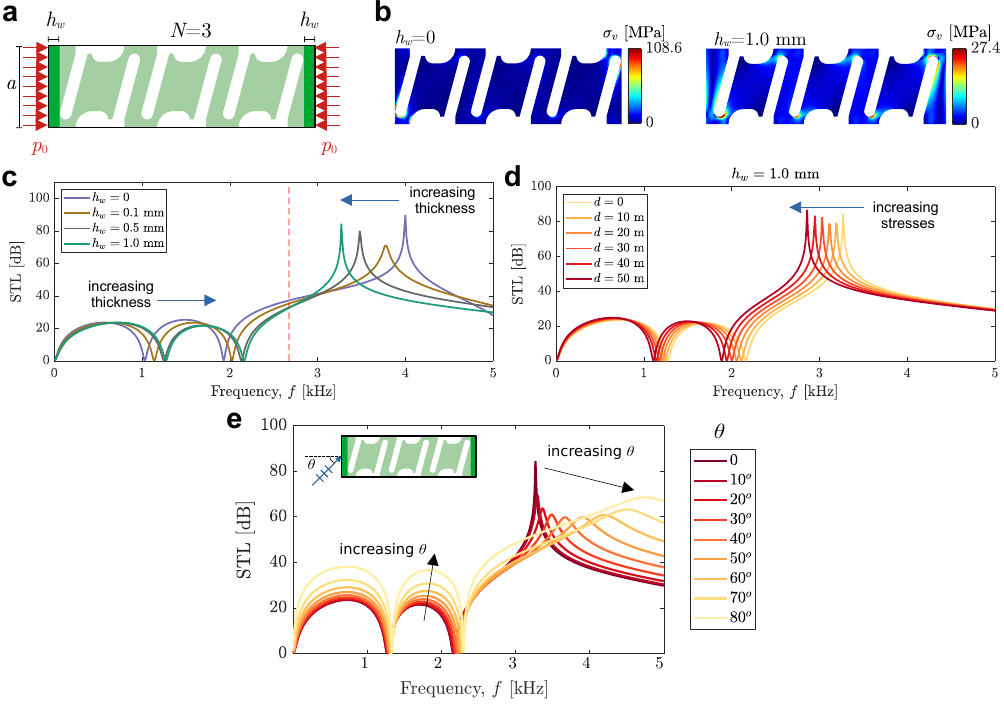}
 \caption{
 Effects of hydrostatic pressure on the stress distribution and STL curves.
 (\textbf{a}) Finite structure obtained by adding two layers with a thickness $h_w$ made of a homogeneous isotropic material to the previously obtained unit cell configuration, chosen with $N=3$. This structure is submerged into water, undergoing a hydrostatic pressure $p_0$ on its lateral walls.
 (\textbf{b}) The value of $h_w$ is increased from $h_w=0$ to $h_w=1.0$ mm, indicating the decrease of the maximum von Mises stress from $108.6$ (top panel) to $27.4$ MPa (bottom panel).
 (\textbf{c}) STL curves for increasing values of $h_w$, indicating a blueshift (redshift) in the frequency range below (above) $2.7$ kHz.
 This difference in behaviour is owed to the diverse attenuation formation mechanisms in each frequency range.
 (\textbf{d}) STL curves for the case $h_w=1.0$ mm for increasing depth values $d = \{0,10,20,30,40,50\}$ m, indicating a redshift in the STL curves due to the development of compressive stresses.
 (\textbf{e}) STL variation with respect to increasing incidence angle ($\theta$), showing an improvement in performance in the low-frequency range and a red-shift of the highest STL peak for increasing incidence angles, followed by a decrease attenuation.
 }
 \label{figure5}
\end{figure}

An immediate concern considering the applicability of the metabarrier is the stress levels that might arise due to the hydrostatic pressure after its submersion in water. 
In particular, prohibitive stress concentrations might occur due to the geometrical variations presented by the optimized unit cell.
To this end, we propose a solution which consists in increasing the unit cell overall thickness by adding layers of a homogeneous material with thickness $h_w$ to a desired number of unit cells (Fig.~\ref{figure5}\textbf{a}), resulting in a total length $h = 3a + 2h_w$.

To determine a suitable value of $h_w$, we perform a quasi-static analysis by imposing a pressure at the lateral walls of the metabarrier given by $p_0 = \rho_w g d$, where $g=9.81$ m/s$^2$ is the gravitational acceleration and $d$ is the depth in which the structure is submerged.
The bottom and top edges of this unit cell present an anti-symmetric condition (opposing displacements).
A non-linear FE analysis is performed considering large displacements with a Total Lagrangian formulation~\cite{bathe2006finite}, since \textit{a priori} linearity is not guaranteed and geometrical buckling might occur.
We then evaluate the stress distribution in the structure for the maximum depth of $d=50$ m~\cite{vasconcelos2024metamaterial} (i.e., $p_0 = 0.49$ MPa). For the initial case $h_w=0$ (i.e., no additional thickness), the maximum computed von Mises stress is $108.6$ MPa (Fig.~\ref{figure5}\textbf{b}, left panel). We then increase the thickness $h_w$ until this maximum von Mises stress reaches a value close to $50\%$ of the material's tensile strength ($55$ MPa, no yield strength available from the manufacturer's datasheet). We then end up with a final thickness of $h_w=1.0$ mm for a maximum von Mises stress of $27.4$ MPa (Fig.~\ref{figure5}\textbf{b}, right panel).
We also verify that the maximum von Mises stress is indeed highly concentrated, while the overall mean von Mises stress levels are considerably lower.
It is also worth to notice that the maximum lateral displacements at the left and right edges, computed for the values $h_w = \{ 0, 0.1, 0.5, 1.0 \}$ mm, are negligible with respect to the unit cell total length, presenting, respectively, the values $0.48$, $0.38$, $0.22$, and $0.03$ mm.


Next, we verify the influence of the additional thickness on the STL curves by re-computing these for increasing values of $h_w$ with $d=0$ (i.e., only the influence of $h_w$ is assessed at these analyses).
In this case, we still consider the $y$-direction lattice length to be equal to $a$, since the maximum vertical displacements at the bottom and top edges of the metabarrier unit cell are equal to $94.5$, $82.8$, $49.9$, and $6.2$ $\mu$m, respectively.
The results, shown in Fig.~\ref{figure5}\textbf{c}, show that increasing the lateral thickness $h_w$ has a two-fold effect on the STL curves, which present (i) a blue-shift effect in the region below $2.67$ kHz and (ii) a red-shift effect in the region above $2.67$ kHz. 
This observation is in agreement with the previous statement that the formation mechanism for the STL peaks and dips is different for the regions divided by the frequency corresponding to the wave mode formed at the X-point (see Fig.~\ref{figure3}\textbf{c}), namely, Bragg-scattering-like (higher frequencies) and due to the effective medium properties (lower frequencies). 
This is also verified by re-computing the effective constitutive matrix for each case, from which we obtain the values of $\delta$ equal to $0.9851$, $0.9848$, and $0.9847$, respectively, for $h_w$ equal to $0.1$, $0.5$, and $1.0$ mm. Such decrease in the value of $\delta$ explains both the red-shift and the slight decrease in the STL peaks below $2.67$ kHz.
At the low-frequency region it is also possible to verify that the behavior of the cases $0.5$ and $1.0$ mm are similar, which is explained by their approximate $\delta$ value.
On the other hand, the STL curves obtained at higher frequencies present the opposed behavior, which however cannot be directly correlated to the band diagram presented in Fig.~\ref{figure3}\textbf{c} due to the broken periodicity of the system caused by the introduction of the additional thickness.
We also note that the STL peaks concerning the case $h_w = 1.0$ mm now present the values $23.5$, $21.5$, and $84.3$ dB, respectively, at $0.73$, $1.71$, and $3.27$ kHz.

The last investigation we perform concerning this metabarrier unit cell design is the influence of the hydrostatic pressure on the STL curves.
To this end, we consider the inclusion of an additional stiffness matrix~\cite{cook2007concepts} caused by the stresses induced by hydrostatic pressure. Figure~\ref{figure5}\textbf{d} presents this variation with respect to the values $d = \{0, 10, 20, 30, 40, 50\}$ m. It is possible to observe a red-shift effect concerning the whole STL curve. This effect is explained by the compressive stresses induced in the structure, which cause an overall softening of the metabarrier, thus decreasing its characteristic frequencies. We note that for the case $d=50$ m, the highest peak is shifted down to $2.86$ kHz ($12.5\%$ reduction with respect to $3.27$ kHz for $d=0$), presenting a mild variation in the peak value (smaller than $2.7 \%$).

Finally, the performance of the metabarrier with respect to varying incidence angles is computed. The results, shown in Fig.~\ref{figure5}\textbf{e}, indicate that in the first frequency region (below $2.67$ kHz), the STL curves present a mild red-shift, especially at higher frequencies (close to $2$ kHz), however presenting also a significant increase in attenuation for larger incidence angles (close to $40$ dB at $\theta = 80^o$). Also, for the second frequency region (above $2.67$ kHz), the STL peak presents a decreased attenuation and a much more significant red-shift effect, however remaining above $60$ dB for all computed cases. Once again, the different behaviors displayed by the structure when considering these frequency ranges are due to the different mechanisms that yield attenuation in each case.

\clearpage
\subsection{Finite two-dimensional circular structure}

\begin{figure}[h]
 \centering
 \includegraphics[scale=1]{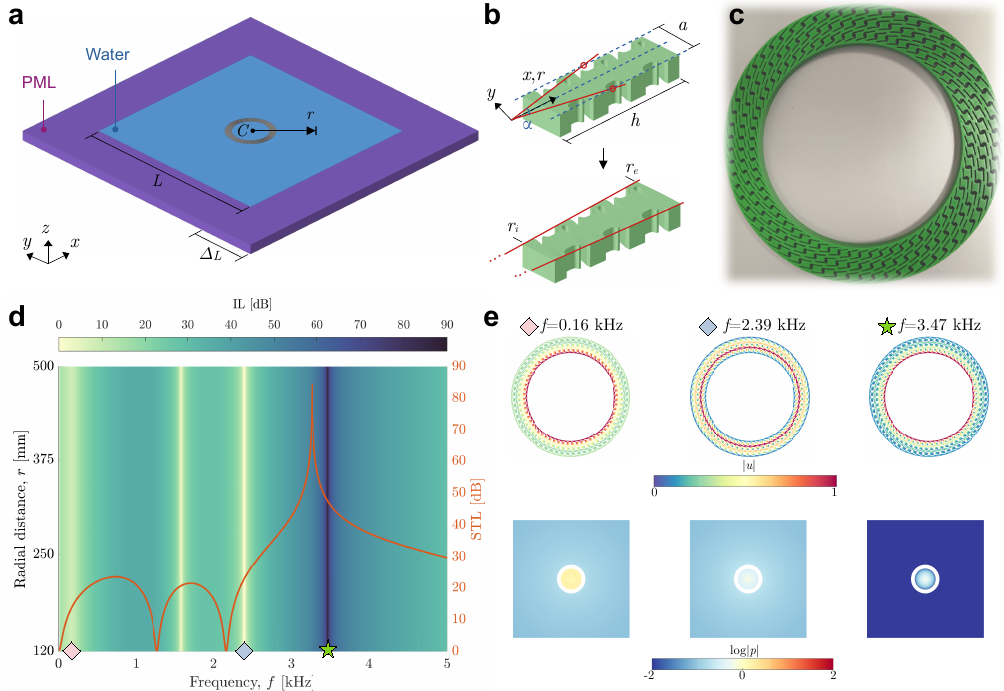}
 \caption{
 (\textbf{a}) Two-dimensional square medium with side length $L$, where a curved metabarrier, centered around $C$, is immersed in water. The radial distance $r$ is measured with respect to $C$.
 An additional PML layer (side length $\Delta L$) is included in the external boundary of the fluid medium.
 (\textbf{b}) The $x$- ($y$-)coordinates of the initially rectangular unit cell (top  panel) are used to generate corresponding radial $r$ (angular $\alpha$) coordinates and obtain a section of a circular shape, yielding an internal radius $r_i$ and an external radius $r_e$ (bottom panel).
 (\textbf{c}) Manufactured structure with $r_i = 88$ mm and $r_e = 120$ mm.
 (\textbf{d}) IL computed for an input at point $C$, varying with respect to the radial distance $r$ and frequency $f$.
 The STL curve obtained for the rectangular unit cell is also shown, for comparison, in red.
 Frequencies corresponding to selected dips are marked using pink and blue diamonds ($0.16$ and $2.39$ kHz, respectively), while a selected peak is marked using a green star ($3.47$ kHz).
 (\textbf{e}) Normalized displacement profiles (top panel) and pressure distribution (bottom panel) for the frequencies indicated in (\textbf{d}). 
 IL dips are associated with non-zero pressure decreases in the vicinity of the metabarrier, while IL peaks lead to significant pressure decrease outside the metabarrier.
 }
 \label{figure6}
\end{figure}

To validate our findings, we now verify the proposed unit cell's performance in protecting an underwater region from noise generated by a point source.
To this end, we dispose the unit cell in a circular manner, enclosing a center point $C$ describing a radial distance $r$ in a two-dimensional square space of length $L$, as shown in Fig.~\ref{figure6}\textbf{a}.
An additional perfectly matched layer (PML) of length $\Delta_L$ is included to minimize reflections from the system boundaries, thus approximating an infinite medium in the $xy$ plane.
The considered medium is square due to the chosen PML FE implementation~\cite{bermudez2006optimal}.

In order to comply with the curved geometry, the metabarrier unit cell must be slightly modified with respect to its original (rectangular) design.
Hence, we position the left (right) edge of the unit cell at $x=r_i$ ($x=r_e$) and its $x$-centerline aligned with $y=0$.
We then compute the angle $\alpha$ described by $C$ with respect to a pair of points located at the center of the bottom ($y = -a/2$) and top edges ($y = +a/2$) of the unit cell, whose $x$ coordinate is given by $x=(r_i+r_e)/2$, as depicted in Fig.~\ref{figure6}\textbf{b}.
This angle is thus defined as $\alpha = 2\tan^{-1}(a/(r_i+r_e))$, yielding a total number of $n_c = 2\pi/\alpha$, $n_c \in \mathcal{N}^*$, unit cells in the angular direction, creating a structure with cyclic symmetry.

Due to fabrication constraints, we set the outer radius of the structure as $r_e = 120$ mm. As the thickness $h=32$ mm of the metabarrier is fixed from the previous design, its inner radius is thus set as $r_i = r_e - h = 88$ mm. From these quantities, we compute $\alpha = 0.0961$ rad, leading to a total number of $n_c = 65.4$ unit cells, which cannot be implemented (non-integer number). We therefore approximate this number to $n_c = 64$ unit cells to ensure a $\pi/2$-rotational symmetry on the metabarrier (to alleviate directional biasing associated with meshing), thus yielding a corrected angle $\alpha = 2\pi/64 = 0.0982$ rad ($\approx 2\%$ deviation). The initial geometry of the unit cell, described in the $(x,y)$ coordinates, is then mapped into a polar coordinate system $(r,\theta)$ following $(x \to r, y \to \theta)$, thus obtaining the final configuration $(x^\prime = r\cos \theta, y^\prime = r \sin \theta)$. The manufactured structure is presented in Fig.~\ref{figure6}\textbf{c}.

We then proceed to perform a FE analysis of the resulting metabarrier embedded in an underwater medium with $L=1000$ mm, considering a PML with length $\Delta L = 250$ mm.
The structural response of the metabarrier and the pressure distribution in the cavity (region encircled by the metabarrier) and external domain are computed using the FE model described in~\cite{mencik2007wave}.
The pressure distribution in the fluid, $p(r,\omega)$, $r \in [r_e,L/2]$, is obtained by imposing a unit acoustic force (representing a monopole source) at the center point $C$, while zero Dirichlet boundary conditions are assigned to the outer edges of the PML (required for numerical accuracy).
In this case, standing waves are formed inside the cavity, which does not allow to separate incident and reflected waves to compute the STL. Instead, we compute the insertion loss (IL) associated with the inclusion of the metabarrier as
\begin{equation}
    \text{IL} = 20 \log \bigg| \frac{\tilde{p}(r,\omega)}{p(r,\omega)} \bigg| \, ,
\end{equation}
where $\tilde{p}(r,\omega)$ is the pressure distribution obtained without the metabarrier.
These results are represented in Fig.~\ref{figure6}\textbf{d}, where $r$ is taken as directed along the $x$-axis of the two-dimensional domain, for simplicity.
Also, for the sake of comparison, we overlay the STL curve (in red) obtained for the rectangular unit cell at a normal incidence angle (see Fig.~\ref{figure5}).


These results show that the IL is significantly more influenced by the excitation frequency $\omega$ than by the distance $r$.
Also, although there is an excellent agreement between the IL (colored surface) and the STL curve (red line), two differences can be immediately noticed upon their comparison, which is illustrated comparing these curves for $r=r_e$.
First, an overall blue-shift of the rectangular STL is noticed, being less noticeable for increasing frequencies.
We note, for instance, that the first two dips (with non-zero frequencies) shift from $1.27$ to $1.58$ kHz and from $2.16$ kHz to $2.39$ kHz, respectively, representing increases of $24\%$ and $11\%$ in frequency.
A similar behavior is observed for the peaks, which, for the case of the highest peak, shifts from $3.27$ to $3.47$ kHz, representing an increase of $6\%$ in frequency.
Second, the zero STL previously computed close to the zero frequency is now lifted, generating a IL of $14.4$ dB at $0.01$ kHz, with a subsequent dip of $8.7$ dB at $0.16$ kHz.
Finally, we also note a significant increase in the attenuation for some of the computed STL peaks. The first and second peaks (previously indicated in Fig.~\ref{figure5}\textbf{c}) now present attenuation levels of $36.2$ and $28.4$ dB, respectively representing increases of $54\%$ and $32\%$.
The third peak has a practically constant value of $84.0$ dB (less than $0.4\%$ reduction).

We also compute the displacement and pressure fields at the frequencies $0.16$, $2.39$, and $3.47$ kHz, corresponding to the first dip, second dip, and third peak, as shown in Fig.~\ref{figure6}\textbf{d}, respectively, using a pink diamond, blue diamond, and green star.
Fig.~\ref{figure6}\textbf{e} shows the corresponding normalized displacements (top panel) and pressure fields (normalized with respect to point $C$, top panel).
For the non-zero IL dip (pink diamond), it is possible to notice significant displacements at the inner part of the metabarrier, while a mild decrease in pressure occurs between the cavity and the outer region.
For zero-IL dip (blue diamond), absolute displacements present similar values at the inner and outer faces of the metabarrier, leading to a negligible acoustic attenuation.
On the other hand, for the indicated peak (green star), absolute displacements are significantly concentrated at the inner face of the metabarrier, resulting in strong pressure decreases at the outer region.

\clearpage
\section{Conclusions} \label{sec_conclusions}

In conclusion, we have proposed a strategy to exploit structural anisotropy in the design of metamaterial-based acoustic barriers yielding efficient underwater noise reduction.
Using our proposed transfer matrix approach, we have calculated the STL curves considering an anisotropic homogeneous medium immersed in a fluid. The first peak of these curves presents (i) an increasing maximum value and (ii) a decreasing frequency, as a conveniently defined mode coupling factor approaches unity.
We have also shown that the thickness of the anisotropic barrier has a lesser influence on the maximum STL value, suggesting that this can be used to tune the frequency of the STL peaks.
This effect is contrary to Bragg-scattering band gaps, whose efficiency in wave attenuation is proportional to the number of unit cells (and therefore thickness) of the structure.

These results were then used to propose a topology optimization problem with the objective of maximizing the mode coupling factor, approaching it to a unit value.
The dispersion relation obtained using the optimized unit cell indicates a high degree of longitudinal-transverse polarization, which, however, was demonstrated not to be responsible for increasing maximum STL values.
A partial band gap (with respect to one of the propagative branches) is also obtained.
The STL curves computed for an increasing number of unit cells immersed in water confirm the previous conclusions regarding the effect of increasing thickness values on the STL curves.
Interestingly, a sub-wavelength behavior is reported for a single unit cell (thickness-to-wavelength ratio circa $1/70$). 
Furthermore, the effect of the partial band gap is confirmed, achieving almost $90$ dB using three unit cells ($30$ mm). 

The effects of hydrostatic pressure on the configuration of three unit cells require the inclusion of additional layers of homogeneous material ($1$ mm each) on each side, which is sufficient to reduce stress levels by a four-fold.
Including these layers leads to (i) a blue-shift of the STL curves for low frequencies and (ii) a red-shift of the STL curves for higher frequencies, without, however, degrading STL levels.
Likewise, the hydrostatic-induced stresses lead to a red-shift in STL curves due to their compressive nature.
The STL curves also present (i) increases in peaks for low frequencies and (ii) blue-shifts for higher frequencies for increasing incidence angles.

Finally, we proposed a convenient transformation for the unit cell, morphing its shape from a rectangular to a circular design. We have also shown the manufacturability of the structure for an external radius of $120$ mm and thickness $32$ mm. 
The IL curves obtained for this configuration, considering a monopole source at the center of an unbounded domain, confirm the previously computed STL curves, lifting the zero-frequency dip and increasing the attenuation of the STL peaks. 
Thus, not only the proposed structure seems suitable for technological exploitation, but its underlying mechanism of anisotropy (with the associated rational design) also indicates a fruitful direction for further investigation in acoustics. 

\section*{CRediT authorship contribution statement}

\noindent
\textbf{VFDP}: Conceptualization, Data curation, Formal analysis, Investigation, Methodology, Software, Visualization, Writing – original draft. \textbf{MM}: Funding acquisition, Project administration, Resources, Supervision, Validation, Writing – review and editing

\section*{Declaration of Competing Interest}

The authors declare that they have no known competing financial interests or personal relationships that could have appeared to influence the work reported in this paper.

\section*{Acknowledgements}

VFDP and MM are supported by the European Union’s Horizon Europe programme in the framework of the ERC StG POSEIDON under Grant Agreement No. 101039576.

\clearpage
\bibliographystyle{elsarticle-num}
\bibliography{biblio}

\begin{thebibliography}{10}
\expandafter\ifx\csname url\endcsname\relax
  \def\url#1{\texttt{#1}}\fi
\expandafter\ifx\csname urlprefix\endcsname\relax\def\urlprefix{URL }\fi
\expandafter\ifx\csname href\endcsname\relax
  \def\href#1#2{#2} \def\path#1{#1}\fi

\bibitem{duarte2021soundscape}
C.~M. Duarte, L.~Chapuis, S.~P. Collin, D.~P. Costa, R.~P. Devassy, V.~M.
  Eguiluz, C.~Erbe, T.~A.~C. Gordon, B.~S. Halpern, H.~R. Harding, M.~N.
  Havlik, M.~Meekan, N.~D. Merchant, J.~L. Miksis-Olds, M.~Parsons,
  M.~Predragovic, A.~N. Radford, C.~A. Radford, S.~D. Simpson, H.~Slabbekoorn,
  E.~Staaterman, I.~C. Van~Opzeeland, J.~Winderen, X.~Zhang, F.~Juanes, The
  soundscape of the {Anthropocene} ocean, Science 371~(6529) (2021) eaba4658.

\bibitem{national2003ocean}
N.~R. Council, D.~on~Earth, L.~Studies, O.~S. Board, C.~on~Potential Impacts of
  Ambient Noise in the Ocean~on Marine~Mammals,
  \href{https://books.google.fr/books?id=OlupYZ1F3_oC}{Ocean Noise and Marine
  Mammals}, National Academies Press, 2003.
\newline\urlprefix\url{https://books.google.fr/books?id=OlupYZ1F3_oC}

\bibitem{ross2013mechanics}
D.~Ross, Mechanics of underwater noise, Elsevier, 2013.

\bibitem{dong2023underwater}
E.~Dong, P.~Cao, J.~Zhang, S.~Zhang, N.~X. Fang, Y.~Zhang, Underwater acoustic
  metamaterials, National Science Review 10~(6) (2023) nwac246.

\bibitem{dong2020review}
J.~Dong, P.~Tian, Review of underwater sound absorption materials, IOP
  Conference Series: Earth and Environmental Science 508~(1) (2020) 012182.

\bibitem{qu2022underwater}
S.~Qu, N.~Gao, A.~Tinel, B.~Morvan, V.~Romero-Garc{\'\i}a, J.-P. Groby,
  P.~Sheng, Underwater metamaterial absorber with impedance-matched composite,
  Science Advances 8~(20) (2022) eabm4206.

\bibitem{gao2022acoustic}
N.~Gao, Z.~Zhang, J.~Deng, X.~Guo, B.~Cheng, H.~Hou, Acoustic metamaterials for
  noise reduction: a review, Advanced Materials Technologies 7~(6) (2022)
  2100698.

\bibitem{croenne2025review}
C.~Cro{\"e}nne, J.~O. Vasseur, L.~Roux, C.~Audoly, A.-C. Hladky, A review of
  acoustic metamaterials for naval and underwater defense applications: from
  historical concepts to new trends, Acta Acustica 9 (2025) 24.

\bibitem{kuhl1950sound}
W.~Kuhl, Sound Absorption and Sound Absorbers in Water. (Dynamic Properties of
  Rubber and Rubberlike Substances in the Acoustic Frequency Region),
  Department of the Navy, Bureau of Ships, 1950.

\bibitem{hladky1991analysis}
A.-C. Hladky-Hennion, J.-N. Decarpigny, Analysis of the scattering of a plane
  acoustic wave by a doubly periodic structure using the finite element method:
  Application to alberich anechoic coatings, The Journal of the Acoustical
  Society of America 90~(6) (1991) 3356--3367.

\bibitem{fu2021review}
Y.~Fu, I.~I. Kabir, G.~H. Yeoh, Z.~Peng, A review on polymer-based materials
  for underwater sound absorption, Polymer Testing 96 (2021) 107115.

\bibitem{lin2022sound}
C.~Lin, G.~S. Sharma, D.~Eggler, L.~Maxit, A.~Skvortsov, I.~MacGillivray,
  N.~Kessissoglou, Sound radiation from a cylindrical shell with a multilayered
  resonant coating, International Journal of Mechanical Sciences 232 (2022)
  107479.

\bibitem{ivansson2008numerical}
S.~M. Ivansson, Numerical design of alberich anechoic coatings with
  superellipsoidal cavities of mixed sizes, The Journal of the Acoustical
  Society of America 124~(4) (2008) 1974--1984.

\bibitem{romero2020design}
V.~Romero-Garc{\'\i}a, N.~Jimenez, G.~Theocharis, V.~Achilleos, A.~Merkel,
  O.~Richoux, V.~Tournat, J.-P. Groby, V.~Pagneux, Design of acoustic
  metamaterials made of {Helmholtz} resonators for perfect absorption by using
  the complex frequency plane, Comptes Rendus. Physique 21~(7-8) (2020)
  713--749.

\bibitem{duan2021tunable}
M.~Duan, C.~Yu, F.~Xin, T.~J. Lu, Tunable underwater acoustic metamaterials via
  quasi-{Helmholtz} resonance: From low-frequency to ultra-broadband, Applied
  Physics Letters 118~(7) (2021).

\bibitem{duan2023deep}
M.~Duan, C.~Yu, F.~Xin, T.~J. Lu, Deep subwavelength hybrid metamaterial for
  low-frequency underwater sound absorption by quasi-helmholtz resonance, AIP
  Advances 13~(2) (2023).

\bibitem{zhang2018subwavelength}
Y.~Zhang, J.~Pan, K.~Chen, J.~Zhong, Subwavelength and quasi-perfect underwater
  sound absorber for multiple and broad frequency bands, The Journal of the
  Acoustical Society of America 144~(2) (2018) 648--659.

\bibitem{zhao2014backing}
H.~Zhao, J.~Wen, H.~Yang, L.~Lv, X.~Wen, Backing effects on the underwater
  acoustic absorption of a viscoelastic slab with locally resonant scatterers,
  Applied Acoustics 76 (2014) 48--51.

\bibitem{elmer2014new}
K.-H. Elmer, J.~Savery, New hydro sound dampers to reduce piling underwater
  noise, INTER-NOISE and NOISE-CON Congress and Conference Proceedings 249~(2)
  (2014) 5551--5560.

\bibitem{elzinga2019manuscript}
J.~Elzinga, A.~Mesu, E.~van Eekelen, M.~Wochner, E.~Jansen, M.~Nijhof,
  Installing offshore wind turbine foundations quieter: A performance overview
  of the first full-scale demonstration of the {AdBm} underwater noise
  abatement system, in: Offshore Technology Conference, OTC, 2019, p.
  D021S019R003.

\bibitem{peng2018modelling}
Y.~Peng, A.~Tsouvalas, A.~Metrikine, E.~Belderbos, Modelling and development of
  a resonator-based noise mitigation system for offshore pile driving, in:
  Proceedings of the 25th International Congress on Sound and Vibration, 2018.

\bibitem{wochner2016underwater}
M.~Wochner, K.~Lee, A.~McNeese, P.~Wilson, Underwater noise mitigation from
  pile driving using a tuneable resonator system, in: Proc. 22nd International
  Congress on Acoustics ICA, Buenos Aires, Argentina, 2016.

\bibitem{su2023finite}
S.~Su, S.~Wu, Finite element analysis of acoustic performance of water-filled
  {Helmholtz} resonator with the effect of elastic cavity walls, Journal of
  Physics: Conference Series 2458~(1) (2023) 012017.

\bibitem{lucke2011use}
K.~Lucke, P.~A. Lepper, M.-A. Blanchet, U.~Siebert, The use of an air bubble
  curtain to reduce the received sound levels for harbor porpoises (phocoena
  phocoena), The Journal of the Acoustical Society of America 130~(5) (2011)
  3406--3412.

\bibitem{wursig2000development}
B.~W{\"u}rsig, C.~Greene~Jr, T.~Jefferson, Development of an air bubble curtain
  to reduce underwater noise of percussive piling, Marine environmental
  research 49~(1) (2000) 79--93.

\bibitem{tsouvalas2020underwater}
A.~Tsouvalas, Underwater noise emission due to offshore pile installation: A
  review, Energies 13~(12) (2020) 3037.

\bibitem{brunet2015soft}
T.~Brunet, A.~Merlin, B.~Mascaro, K.~Zimny, J.~Leng, O.~Poncelet,
  C.~Arist{\'e}gui, O.~Mondain-Monval, Soft 3d acoustic metamaterial with
  negative index, Nature materials 14~(4) (2015) 384--388.

\bibitem{tallon2021experimental}
B.~Tallon, A.~Kovalenko, O.~Poncelet, C.~Arist{\'e}gui, O.~Mondain-Monval,
  T.~Brunet, Experimental demonstration of negative refraction with 3d locally
  resonant acoustic metafluids, Scientific Reports 11~(1) (2021) 4627.

\bibitem{pierre2014resonant}
J.~Pierre, B.~Dollet, V.~Leroy, Resonant acoustic propagation and negative
  density in liquid foams, Physical review letters 112~(14) (2014) 148307.

\bibitem{pierre2017investigating}
J.~Pierre, C.~Gaulon, C.~Derec, F.~Elias, V.~Leroy, Investigating the origin of
  acoustic attenuation in liquid foams, The European Physical Journal E 40
  (2017) 1--11.

\bibitem{jansen2012experimental}
H.~Jansen, C.~de~Jong, B.~Jung, Experimental assessment of the insertion loss
  of an underwater noise mitigation screen for marine pile driving, in:
  Proceedings of the 11th European Conference on Underwater Acoustics-ECUA
  2012, 2-6 July 2012, Edinburgh, UK, 2012.

\bibitem{koschinski2013development}
S.~Koschinski, K.~L{\"u}demann, Development of noise mitigation measures in
  offshore wind farm construction, Commissioned by the Federal Agency for
  Nature Conservation (2013) 1--102.

\bibitem{wang2022topological}
Y.~Wang, H.~Zhao, H.~Yang, J.~Liu, D.~Yu, J.~Wen, Topological design of lattice
  materials with application to underwater sound insulation, Mechanical Systems
  and Signal Processing 171 (2022) 108911.

\bibitem{chen2020highly}
Y.~Chen, B.~Zhao, X.~Liu, G.~Hu, Highly anisotropic hexagonal lattice material
  for low frequency water sound insulation, Extreme Mechanics Letters 40 (2020)
  100916.

\bibitem{wang2023acoustically}
Y.~Wang, H.~Zhao, H.~Yang, H.~Zhang, T.~Li, C.~Wang, J.~Liu, J.~Zhong, D.~Yu,
  J.~Wen, Acoustically soft and mechanically robust hierarchical metamaterials
  in water, Physical Review Applied 20~(5) (2023) 054015.

\bibitem{wang2024topology}
C.~Wang, H.~Zhao, Y.~Wang, J.~Zhong, D.~Yu, J.~Wen, Topology optimization of
  chiral metamaterials with application to underwater sound insulation, Applied
  Mathematics and Mechanics 45~(7) (2024) 1119--1138.

\bibitem{lai2009introduction}
W.~M. Lai, D.~Rubin, E.~Krempl, Introduction to Continuum Mechanics,
  Butterworth-Heinemann, 2009.

\bibitem{jimenez2021acoustic}
N.~Jim{\'e}nez, O.~Umnova, J.-P. Groby, Acoustic waves in periodic structures,
  metamaterials, and porous media, Springer, 2021.

\bibitem{oudich2014general}
M.~Oudich, X.~Zhou, M.~Badreddine~Assouar, General analytical approach for
  sound transmission loss analysis through a thick metamaterial plate, Journal
  of Applied Physics 116~(19) (2014).

\bibitem{kinsler2000fundamentals}
L.~Kinsler, A.~Frey, A.~Coppens, J.~Sanders,
  \href{https://books.google.fr/books?id=FecSEAAAQBAJ}{Fundamentals of
  Acoustics}, Wiley, 2000.
\newline\urlprefix\url{https://books.google.fr/books?id=FecSEAAAQBAJ}

\bibitem{kweun2017transmodal}
J.~M. Kweun, H.~J. Lee, J.~H. Oh, H.~M. Seung, Y.~Y. Kim, Transmodal
  {Fabry-P{\'e}rot} resonance: theory and realization with elastic
  metamaterials, Physical review letters 118~(20) (2017) 205901.

\bibitem{cook2007concepts}
R.~D. Cook, Concepts and applications of finite element analysis, John Wiley \&
  Sons, 2007.

\bibitem{pinho2009asymptotic}
J.~Pinho-da Cruz, J.~Oliveira, F.~Teixeira-Dias, Asymptotic homogenisation in
  linear elasticity. part i: Mathematical formulation and finite element
  modelling, Computational Materials Science 45~(4) (2009) 1073--1080.

\bibitem{goldberg2013genetic}
D.~Goldberg, \href{https://books.google.fr/books?id=6gzS07Sv9hoC}{Genetic
  Algorithms}, Pearson Education India, 2013.
\newline\urlprefix\url{https://books.google.fr/books?id=6gzS07Sv9hoC}

\bibitem{mace2008modelling}
B.~R. Mace, E.~Manconi, Modelling wave propagation in two-dimensional
  structures using finite element analysis, Journal of Sound and Vibration
  318~(4-5) (2008) 884--902.

\bibitem{bathe2006finite}
K.-J. Bathe, Finite element procedures, Klaus-Jurgen Bathe, 2006.

\bibitem{vasconcelos2024metamaterial}
A.~C.~A. Vasconcelos, S.~V. Valappil, D.~Schott, J.~Jovanova, A.~M. Arag{\'o}n,
  A metamaterial-based interface for the structural resonance shielding of
  impact-driven offshore monopiles, Engineering Structures 300 (2024) 117261.

\bibitem{bermudez2006optimal}
A.~Berm{\'u}dez, L.~Hervella-Nieto, A.~Prieto, R.~Rodr{\i}guez, An optimal
  finite-element/{PML} method for the simulation of acoustic wave propagation
  phenomena, Variational Formulations in Mechanics: Theory and
  Applications,(January) (2006).

\bibitem{mencik2007wave}
J.-M. Mencik, M.~Ichchou, Wave finite elements in guided elastodynamics with
  internal fluid, International Journal of Solids and Structures 44~(7-8)
  (2007) 2148--2167.

\bibitem{slaughter2012linearized}
W.~S. Slaughter, The linearized theory of elasticity, Springer Science \&
  Business Media, 2012.

\bibitem{bloch1929quantenmechanik}
F.~Bloch, {\"U}ber die {Quantenmechanik} der {Elektronen} in {Kristallgittern},
  Zeitschrift f{\"u}r {Physik} 52~(7-8) (1929) 555--600.

\bibitem{yang2017prediction}
Y.~Yang, B.~R. Mace, M.~J. Kingan, Prediction of sound transmission through,
  and radiation from, panels using a wave and finite element method, The
  Journal of the Acoustical Society of America 141~(4) (2017) 2452--2460.

\end{thebibliography}

\clearpage
\appendix
\setcounter{figure}{0}

\section{Constitutive equations in a general anisotropic two-dimensional medium} \label{sec_constitutive}

The terms of the constitutive matrix can be expressed with respect to general coordinates $x^\prime y^\prime z$, rotated with respect to the original $xyz$ coordinates by an angle $\beta$ around the $z$-axis using the relation
\begin{equation}
    \mathbf{C}^\prime = \mathbf{T}^T \mathbf{C} \mathbf{T} \, ,
\end{equation}
where $\mathbf{T}$ is a coordinate transformation matrix, expressed by~\cite{cook2007concepts}
\begin{equation}
    \mathbf{T} = \left[\
        \begin{array}{ccc}
            \cos^2 \beta & \sin^2 \beta & \sin \beta \cos \beta  \\
            \sin^2 \beta & \cos^2 \beta & -\sin \beta \cos \beta  \\
            -2 \sin \beta \cos \beta & 2 \sin \beta \cos \beta & \cos^2\beta - \sin^2 \beta \\
        \end{array}
    \right] \, .
\end{equation}

In the case of an isotropic two-dimensional medium, $C_{13} = C_{31} = C_{23} = C_{32} = 0$, and the ratio between the remaining constants is such that $C^\prime_{13} = C^\prime_{31} = C^\prime_{23} = C^\prime_{32} = 0$, as well, for any angle $\beta$. On the other hand, anisotropic media may present $C_{13} = C_{31} \neq 0$ and $C_{23} = C_{32} \neq 0$, thus evidencing the coupling between normal stresses and shear strains (and vice-versa).

\section{Wave propagation in a homogeneous anisotropic medium}
\label{sec_model_wave_homogeneous}

The equations of motion for a two-dimensional medium free of external forces can be written in tensor notation as~\cite{slaughter2012linearized}
\begin{equation} \label{motion1}
    \sigma_{ji,j} = \rho \ddot{u}_i \, ,
\end{equation}
where the partial derivatives with respect to the spatial coordinate $j$ and time coordinate $t$ are denoted, respectively, as $(\circ)_{,j} = \frac{\partial (\circ)}{\partial j}$ and $(\ddot{\circ}) = \frac{\partial^2 (\circ)}{\partial t^2}$, and $\rho$ is the material specific mass density.

Equations (\ref{constitutive}) and (\ref{motion1}) can be combined and rewritten by considering an infinitesimal strain tensor ($\varepsilon_{kl} = \varepsilon_{lk} = \frac{1}{2}( u_{k,l} + u_{l,k} )$), yielding
\begin{equation} \label{motion2}
 \mathbf{C}_x \mathbf{u}_{,xx} +
 \mathbf{C}_{xy} \mathbf{u}_{,xy} +
 \mathbf{C}_y \mathbf{u}_{,yy} =
 \rho \ddot{\mathbf{u}} ,
\end{equation}
where the matrix and vector components are described as
\begin{equation}
 \mathbf{C}_x = \left[
 \begin{array}{cc}
  C_{11} & C_{13} \\
  C_{31} & C_{33} 
 \end{array}
 \right], \,
 \mathbf{C}_{xy} = \left[
 \begin{array}{cc}
  C_{13}+C_{31} & C_{12}+C_{33}  \\
  C_{21}+C_{33} & C_{23}+C_{32}
 \end{array}
 \right], \,
 \mathbf{C}_y = \left[
 \begin{array}{cc}
  C_{33} & C_{32} \\
  C_{23} & C_{22}
 \end{array}
 \right], \,
 \mathbf{u} = \left\{
 \begin{array}{c}
  u_x \\
  u_y
 \end{array}
 \right\} .
\end{equation}

Consider now a Bloch solution for the displacements in the form \cite{bloch1929quantenmechanik}
\begin{equation} \label{displacement_form}
 \mathbf{u}(\mathbf{r},t) = \mathbf{U} e^{-\text{i} \omega t} e^{\text{i} \mathbf{k} \cdot \mathbf{r} },
\end{equation}
where $\mathbf{U} = \{ U_x, U_y \}^T$ is a wave mode, 
$\omega$ is the circular frequency,
$\mathbf{k} = \{ k_x, k_y \}^T$ is the two-dimensional wave vector, and
$\mathbf{r} = \{ x, y \}^T$ is the two-dimensional coordinate vector.
Substituting Eq. (\ref{displacement_form}) in (\ref{motion2}) yields
\begin{equation}
    (k_x^2 \mathbf{C}_x + k_x k_y \mathbf{C}_{xy} + k_y^2 \mathbf{C}_y - \omega^2 \rho \mathbf{I} ) \mathbf{U} = \mathbf{0} \, ,
\end{equation}
where $\mathbf{I}$ a order-$2$ identity matrix. The previous equation can be solved as a quadratic eigenvalue problem for $k_x$ ($k_y$) for a specified value of $k_y$ ($k_x$).
Rewriting the wavenumber components $k_x = k \cos \theta$ and $k_y = k \sin \theta$, for a given wavenumber $k$ and propagation direction $\theta$, leads to the eigenproblem
\begin{equation} \label{dispersion_bulk}
 ( k^2 \mathbf{D} - \omega^2 \rho \mathbf{I} ) \mathbf{U} = \mathbf{0} \, ,
\end{equation}
where $\mathbf{D} = \mathbf{D}(\mathbf{C},\theta) = \mathbf{C}_x \cos^2 \theta + \mathbf{C}_{xy} \sin \theta \cos \theta + \mathbf{C}_y \sin^2 \theta $ is a direction-dependent stiffness matrix. By considering the symmetry of the constitutive matrix, the terms of $\mathbf{D}$ can be written as
\begin{equation}
    \begin{aligned}
        D_{11} &= C_{11} \cos^2 \theta + C_{13} \sin 2\theta + C_{33} \sin^2 \theta \, , \\
        D_{12} = D_{21} &= C_{13} \cos^2 + \frac{C_{12}+C_{33}}{2} \sin 2\theta + C_{23} \sin^2 \theta \, , \\
        D_{22} &= C_{33} \cos^2 + C_{23} \sin 2\theta + C_{22} \sin^2 \theta \, . \\
    \end{aligned}
\end{equation}

The dispersion relation $k = k(\omega)$ in a homogeneous anisotropic medium can then be obtained for a given propagation direction $\theta$ by computing the determinant of Eq. (\ref{dispersion_bulk}).
Considering a non-singular determinant of matrix $\mathbf{D}$ (i.e., $D_{11}D_{12} - D_{12}D_{21} \neq 0$) yields two pairs of non-dispersive solutions $k(\omega) = \pm k_1(\omega) = \pm \omega/c_1(\omega)$ and $k(\omega) = \pm k_2(\omega) = \pm \omega/c_2(\omega)$, where the wave speeds $c_{1,2}$ are given by
\begin{equation} \label{dispersion_relation}
    c_{1,2}(\mathbf{C},\theta) = \sqrt{ \frac{1}{\rho} \frac{2( D_{11}D_{22}-D_{12}D_{21} )}{ (D_{11}+D_{22}) \pm \sqrt{  (D_{11}-D_{22})^2 + 4 D_{12}D_{21} } }  } \, ,
\end{equation}
and the subscript $1$ ($2$) refers to the solution containing the $+$ ($-$) sign.

This pair of wavenumbers ($k_1$, $k_2$) can be substituted back into Eq. (\ref{dispersion_bulk}) to obtain the normalized mode shapes $ \mathbf{U}_i = \{ \cos \phi_i, \, \sin \phi_i \}^T$, $i = \{1,2\}$, related by
\begin{equation} \label{phi_angles}
 \bigg( \frac{U_y}{U_x} \bigg)_{1,2} = \tan \phi_{1,2} =
 \frac{D_{11} [ (D_{11}+D_{22}) \pm \sqrt{(D_{11}-D_{22})^2 + 4D_{12}D_{21})} ] - 2(D_{11}D_{22} - D_{12}D_{21} ) }{ -D_{12} [ (D_{11}+D_{22}) \pm \sqrt{(D_{11}-D_{22})^2 + 4D_{12}D_{21}} ] } \, .
\end{equation}

For the simplified case of a four-fold symmetric unit cell (i.e., with respect to both $x$ and $y$ axes), one has $C_{13} = C_{31} = 0$; considering waves propagating in the $x$-direction ($\theta=0$), the well-know solutions $k_l(\omega) = \omega \sqrt{ \rho/C_{11} }$ and $k_s(\omega) = \omega \sqrt{ \rho/C_{33} }$ for longitudinal and shear waves, respectively, are retrieved.
The corresponding wave modes, described as $\mathbf{U}_l = \left\{ 1, 0 \right\}^T$ and $\mathbf{U}_s = \left\{ 0, 1 \right\}^T$, indicate a full decoupling between the corresponding motions.
This simplification is no longer valid for $\theta=0$ whenever $C_{13} \neq 0$.

\section{Derivation of transfer matrix method} \label{sec_derivation_tmm}

The shear stresses $\tau_{xy}$ in the metabarrier can be written combining Eqs.~(\ref{constitutive})--(\ref{app_displacements_normal}) as
\begin{equation}
 \tau_{xy}(x) = 
   \hat{u}_{1p} C_z^{\phi_1} \text{i}k_1 e^{\text{i}k_1 x}
 - \hat{u}_{1n} C_z^{\phi_1} \text{i}k_1 e^{-\text{i}k_1 x}
 + \hat{u}_{2p} C_z^{\phi_2} \text{i}k_2 e^{\text{i}k_2 x} 
 - \hat{u}_{2n} C_z^{\phi_2} \text{i}k_2 e^{-\text{i}k_2 x} \, ,
\end{equation}
where $C_z^{\phi} = C_{31} \cos \phi + C_{33} \sin \phi$, with $\phi_i = \angle \mathbf{U}_i$. Time dependencies, including the $e^{-\text{i} \omega t}$ terms, are henceforth omitted for the sake of brevity.

At the fluid-structure interfaces between metabarrier and surrounding fluid, shear stresses vanish~\cite{oudich2014general}, i.e., 
\begin{equation} \label{interface_shear}
 \tau_{xy}(x=0) = 0,
 \tau_{xy}(x=h) = 0,
\end{equation}
which can be substituted into the previous equation, leading to
\begin{equation}
\begin{aligned}
 \left[
    \begin{array}{cc}
     C_z^{\phi_1} k_1 & C_z^{\phi_2} k_2 \\
     C_z^{\phi_1} k_1 e^{\text{i}k_1 h} & C_z^{\phi_2} k_2 e^{\text{i}k_2 h} 
    \end{array}
 \right]
 \left\{
    \begin{array}{c}
     \hat{u}_{1p} \\
     \hat{u}_{2p}
    \end{array}
 \right\} 
 - 
 \left[
    \begin{array}{cc}
     C_z^{\phi_1} k_1 & C_z^{\phi_2} k_2 \\
     C_z^{\phi_1} k_1 e^{-\text{i}k_1 h} & C_z^{\phi_2} k_2 e^{-\text{i}k_2 h}
    \end{array}
 \right]
 \left\{
    \begin{array}{c}
     \hat{u}_{1n} \\
     \hat{u}_{2n}
    \end{array}
 \right\}
 = \mathbf{0} \, ,
\end{aligned}
\end{equation}
thus allowing to obtain a relation between the complex amplitudes of negative- and positive-going waves as
\begin{equation} \label{app_ratio_positive_negative}
 \left\{
    \begin{array}{c}
     \hat{u}_{1n} \\
     \hat{u}_{2n}
    \end{array}
 \right\} =
 \left[
    \begin{array}{cc}
     \Delta_{11} & \Delta_{12} \\
     \Delta_{21} & \Delta_{22}
    \end{array}
 \right]
 \left\{
    \begin{array}{c}
     \hat{u}_{1p} \\
     \hat{u}_{2p}
    \end{array}
 \right\} \, ,
\end{equation}
where
\begin{equation}
\begin{aligned}
 \Delta_{11} &= \frac{e^{-\text{i} k_2 h} - e^{\text{i} k_1 h}}{e^{-\text{i} k_2 h} - e^{-\text{i} k_1 h}} \, , 
 \Delta_{12} = \frac{C_z^{\phi_2} k_2}{C_z^{\phi_1} k_1} \frac{e^{-\text{i} k_2 h} - e^{\text{i} k_2 h}}{e^{-\text{i} k_2 h} - e^{-\text{i} k_1 h}} \, , \\
 \Delta_{21} &= \frac{C_z^{\phi_1} k_1}{C_z^{\phi_2} k_2} \frac{e^{-\text{i} k_1 h} - e^{\text{i} k_1 h}}{e^{-\text{i} k_2 h} - e^{-\text{i} k_1 h}} \, , 
 \Delta_{22} = \frac{e^{\text{i} k_2 h} - e^{-\text{i} k_1 h}}{e^{-\text{i} k_2 h} - e^{-\text{i} k_1 h}} \, .
\end{aligned}
\end{equation}

Next, we apply a similar procedure considering normal stresses, which can be written as
\begin{equation}
 \sigma_x(x) = 
   \hat{u}_{1p} C_x^{\phi_1} \text{i}k_1 e^{\text{i}k_1 x}
 - \hat{u}_{1n} C_x^{\phi_1} \text{i}k_1 e^{-\text{i}k_1 x}
 + \hat{u}_{2p} C_x^{\phi_2} \text{i}k_2 e^{\text{i}k_2 x} 
 - \hat{u}_{2n} C_x^{\phi_2} \text{i}k_2 e^{-\text{i}k_2 x},
\end{equation}
where $C_x^{\phi} = C_{11} \cos \phi + C_{13} \sin \phi$ and the time dependence is once again omitted.

A state vector containing the $x$-direction displacement ($u_x$) and normal stress ($\sigma_x$) can now be evaluated in the $x$-coordinate as
\begin{equation}
 \left\{
    \begin{array}{c}
     u_x \\
     \sigma_x
    \end{array}
 \right\}_x =
 \left[
    \begin{array}{cccc}
     \cos \phi_1 e^{\text{i}k_1x} & \cos \phi_1 e^{-\text{i}k_1x} & \cos \phi_2 e^{\text{i}k_2x} & \cos \phi_2 e^{-\text{i}k_2x} \\
     C_x^{\phi_1} \text{i} k_1 e^{\text{i}k_1x} & -C_x^{\phi_1} \text{i} k_1 e^{-\text{i}k_1x} & C_x^{\phi_2} \text{i} k_2 e^{\text{i}k_2x} & -C_x^{\phi_2} \text{i} k_2 e^{-\text{i}k_2x} \\
    \end{array}
 \right]
 \left\{
    \begin{array}{c}
     \hat{u}_{1p} \\
     \hat{u}_{1n} \\
     \hat{u}_{2p} \\
     \hat{u}_{2n} \\
    \end{array}
 \right\} \, , 
\end{equation}
which can be combined with Eq.~(\ref{app_ratio_positive_negative}) to express the state vector as
\begin{equation}
 \left\{
    \begin{array}{c}
     u_x \\
     \sigma_x
    \end{array}
 \right\}_x =
 \mathbf{M}(x)
  \left\{
    \begin{array}{c}
     \hat{u}_{1p} \\
     \hat{u}_{2p} \\
    \end{array}
 \right\} \, ,
\end{equation}
where
\begin{equation}
 \mathbf{M}(x) = 
 \left[
    \begin{array}{cccc}
     \cos \phi_1 e^{\text{i}k_1x} & \cos \phi_1 e^{-\text{i}k_1x} & \cos \phi_2 e^{\text{i}k_2x} & \cos \phi_2 e^{-\text{i}k_2x} \\
     C_x^{\phi_1} \text{i} k_1 e^{\text{i}k_1x} & -C_x^{\phi_1} \text{i} k_1 e^{-\text{i}k_1x} & C_x^{\phi_2} \text{i} k_2 e^{\text{i}k_2x} & -C_x^{\phi_2} \text{i} k_2 e^{-\text{i}k_2x} \\
    \end{array}
 \right]
 \left[
    \begin{array}{cc}
     1 & 0 \\
     \Delta_{11} & \Delta_{12} \\
     0 & 1 \\
     \Delta_{21} & \Delta_{22} \\
    \end{array}
 \right] \, .
\end{equation}

Combining $\{ u_x, \sigma_x \}_{x=h}^T = \mathbf{M}(x=h) \{ \hat{u}_{1p}, \hat{u}_{2p} \}^T $ and 
$\{ u_x, \sigma_x \}_{x=0}^T = \mathbf{M}(x=0) \{ \hat{u}_{1p}, \hat{u}_{2p} \}^T $ allows to relate the displacements and stresses at the edges of the metabarrier as
\begin{equation} \label{app_transfer_matrix_s}
 \left\{
    \begin{array}{c}
     u_x \\
     \sigma_x
    \end{array}
 \right\}_{x=h}
 = \mathbf{T}^{(s)}
  \left\{
    \begin{array}{c}
     u_x \\
     \sigma_x
    \end{array}
 \right\}_{x=0} 
 = 
 \left[
    \begin{array}{cc}
     T^{(s)}_{11} & T^{(s)}_{12} \\
     T^{(s)}_{21} & T^{(s)}_{22}
    \end{array}
 \right]
 \left\{
    \begin{array}{c}
     u_x \\
     \sigma_x
    \end{array}
 \right\}_{x=0} ,
\end{equation}
where $\mathbf{T}^{(s)} = \mathbf{M}(x=h) \mathbf{M}^{-1}(x=0)$ represents a real-valued structural transfer matrix.
Also, $T_{11}^{(s)} = T_{22}^{(s)}$ and $\det(\mathbf{T}^{(s)}) = T_{11}^{(s)}T_{22}^{(s)} - T_{12}^{(s)}T_{21}^{(s)} = 1$ due to the reciprocity between both propagation directions.
The complete expression of the terms in matrix $\mathbf{T}^{(s)}$ is too long to be conveniently expressed, in which case, a numerical solution is preferred. We now proceed to the coupling between the structure and surrounding fluid.

The coupling between the solid metabarrier and the surrounding fluid can be considered by ensuring acceleration and stress continuity conditions at both edges of the metabarrier. The continuity of accelerations can be stated as~\cite{oudich2014general}
\begin{equation} \label{interface_acceleration}
\begin{array}{ll}
 (P_i + P_r)_{,x} \big|_{x=0} = \rho_0 \omega^2 u_x \big|_{x=0}, &
 (P_t + P_{i^\prime})_{,x} \big|_{x=h} = \rho_0 \omega^2 u_x \big|_{x=h}, 
\end{array}
\end{equation}
and the continuity of stresses at the interface can be written as
\begin{equation} \label{interface_normal_stress}
\begin{array}{ll}
 -(P_i + P_r) \big|_{x=0} = \sigma_x |_{x=0}, &
 -(P_t + P_{i^\prime}) \big|_{x=h} = \sigma_x |_{x=h}.
\end{array}
\end{equation}

Combining Eqs.~(\ref{pressure_waves_x})--(\ref{interface_normal_stress}) yields
\begin{equation} \label{pressure_to_state}
 \left[
    \begin{array}{cc}
     \frac{\text{i}k_0}{\rho_0 \omega^2} & -\frac{\text{i}k_0}{\rho_0 \omega^2} \\
     -1 & -1
    \end{array}
 \right] \mathbf{P} =
 \left\{
    \begin{array}{c}
     u_x \\
     \sigma_x
    \end{array}
 \right\},
\end{equation}
where $\mathbf{P} = \{ \hat{P}_i, \hat{P}_r \}^T$ for $\{ u_x, \sigma_x \}^T_{x=0}$ and
$\mathbf{P} = \{ \hat{P}_t, \hat{P}_{i^\prime} \}^T$ for $\{ u_x, \sigma_x \}_{x=h}$. Combining Eqs.~(\ref{app_transfer_matrix_s}) and (\ref{pressure_to_state}) leads to
\begin{equation}
 \left\{
    \begin{array}{c}
        \hat{P}_t \\
        \hat{P}_{i^\prime} \\
    \end{array}
 \right\}
 =
 \mathbf{T}^{(a)}
 \left\{
    \begin{array}{c}
        \hat{P}_i \\
        \hat{P}_r \\
    \end{array}
 \right\}
 =
 \left[
    \begin{array}{cc}
     T^{(a)}_{11} & T^{(a)}_{12} \\
     T^{(a)}_{21} & T^{(a)}_{22}
    \end{array}
 \right]
 \left\{
    \begin{array}{c}
        \hat{P}_i \\
        \hat{P}_r \\
    \end{array}
 \right\} \, ,
\end{equation}
where the terms of the acoustic transfer matrix $\mathbf{T}^{(a)}$ are given by
\begin{equation}
\begin{aligned}
 T^{(a)}_{11}  = T^{(s)}_{11} + \frac{\text{i}}{2} \bigg(-\frac{k_0}{\rho_0 \omega^2} T^{(s)}_{21} + \frac{\rho_0 \omega^2}{k_0} T^{(s)}_{12} \bigg), \, 
 &T^{(a)}_{12} = \frac{\text{i}}{2} \bigg(\frac{k_0}{\rho_0 \omega^2} T^{(s)}_{21} + \frac{\rho_0 \omega^2}{k_0} T^{(s)}_{12} \bigg), \\ 
 T^{(a)}_{21}  = - \frac{\text{i}}{2} \bigg(\frac{k_0}{\rho_0 \omega^2} T^{(s)}_{21} + \frac{\rho_0 \omega^2}{k_0} T^{(s)}_{12} \bigg), \, 
 &T^{(a)}_{22} = T^{(s)}_{22} + \frac{\text{i}}{2} \bigg(\frac{k_0}{\rho_0 \omega^2} T^{(s)}_{21} - \frac{\rho_0 \omega^2}{k_0} T^{(s)}_{12} \bigg) \, \\ 
\end{aligned}
\end{equation}
and $\det(\mathbf{T}^{(a)}) = T_{11}^{(a)} T_{22}^{(a)} - T_{12}^{(a)}T_{21}^{(a)} = 1$ due to the reciprocity of the system.

\section{Sound transmission loss computation with finite elements}
\label{sec_stl_method_fe}

\begin{figure}[h]
 \centering
 \includegraphics[scale=1]{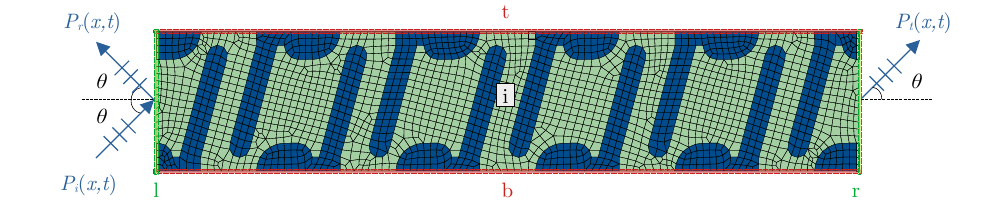}
 \caption{
 Metabarrier STL behaviour under oblique incidence angles and a diffuse field.
 }
 \label{figure_s1}
\end{figure}

The formulation presented in this section is derived from~\cite{yang2017prediction}. The computational approach for the computation of the STL considering a fluid-filled structure is illustrated in Fig.~\ref{figure_s1}, whose FE discretization is shown for illustration purposes. The solid region is indicated by gray elements, while fluid-filled cavities are indicated in blue.

The incident ($P_i$), reflected ($P_r$), transmitted acoustic wave ($P_t$), and FE displacements ($u_q$, $q = \{x,y\}$) are described as
\begin{equation} \label{pu_fourier}
\begin{aligned}
   P_i(x,y,t) &= e^{-\text{i} \omega t} \sum_{G_y} \hat{P}_i(G_y) e^{\text{i} k_x^{(0)} x} e^{\text{i} (k_y^{(0)}+G_y) y} \, , \\
   P_r(x,y,t) &= e^{-\text{i} \omega t} \sum_{G_y} \hat{P}_r(G_y) e^{-\text{i} k_{xa}(G_y) x} e^{\text{i} (k_y^{(0)}+G_y) y} \, , \\
   P_t(x,y,t) &= e^{-\text{i} \omega t} \sum_{G_y} \hat{P}_t(G_y) e^{+\text{i} k_{xa}(G_y) (x-h) } e^{\text{i} (k_y^{(0)}+G_y) y} \, , \\
   u_q        &= e^{-\text{i} \omega t} \sum_{G_y} \hat{u}_q(G_y) e^{\text{i} (k_y^{(0)}+G_y) y} \, ,
\end{aligned}
\end{equation}
where
$G_y = n (2 \pi/a)$, $n \in \mathcal{Z}$, is the reciprocal wavenumber in the $y$-direction for a lattice of length $a$,
$\hat{p}(G_y)$ is the spatial Fourier component corresponding to the quantity $p$ ($P_i$, $P_r$, $P_t$, or $u_q$),
$k_x^{(0)}$ and $k_y^{(0)}$ refer to the wavenumbers in the $x$- and $y$-directions in the fluid, computed as $k_x^{(0)} = k_0 \cos \theta$ and $k_y = k_0 \sin \theta$, where  $k_0 = \omega / c_0$ is the acoustic wavenumber for a sound speed value of $c_0$, 
and $k_xa(G_y)$ is computed such that $k_{xa}^2(G_y) + (k_y + G_y)^2 = k_0^2$.
The shift $(x-h)$ in the exponential term of the transmitted wave is denoted for analytical convenience.

The continuity of accelerations (see Eq.~(\ref{interface_acceleration})) at the incident face (index l in Fig.~\ref{figure_s1}) leads to
\begin{equation} \label{pr_fourier}
    \hat{P}_r(G_y) = \frac{k_x^{(0)}}{k_{xa}(G_y)} \hat{P}_i(G_y) + \text{i} \frac{\rho_e \omega^2}{k_{xa}(G_y)} \hat{u}_x^{\text{(l)}}(G_y) \, ,
\end{equation}
where $\rho_e$ is the specific mass density of the surrounding fluid (outside the finite structure), and the superscript (l) refers to nodes at the incident face (left) of the metabarrier.

Analogously, at the transmitted face, one obtains
\begin{equation} \label{pt_fourier}
    \hat{P}_t(G_y) = - \text{i} \frac{\rho_e \omega^2}{k_{xa}(G_y)} \hat{u}_x^{\text{(r)}}(G_y) \, ,
\end{equation}
where  the superscript (r) refers to nodes at the transmitted face (right) of the metabarrier.


The pressure produced at the input (left) and output (right) faces of the structure can be written, using Eqs.~(\ref{pu_fourier})--(\ref{pt_fourier}), respectively as
\begin{equation} \label{pin_pout}
\begin{aligned}
    P_{\text{in}} = P_i(x=0) + P_r(x=0) &= \sum_{G_y} \bigg[ \bigg( 1 + \frac{k_x^{(0)}}{k_{xa}(G_y)} \bigg) \hat{P}_i(G_y) +  \text{i} \frac{\rho_e \omega^2}{k_{xa}(G_y)} \hat{u}_x^{\text{(l)}}(G_y) \bigg] e^{\text{i} (k_y^{(0)}+G_y) y} \, , \\
    P_{\text{out}} = - P_t(x=h) &= \sum_{G_y} \bigg[ \text{i} \frac{\rho_e \omega^2}{k_{xa}(G_y)} \hat{u}_x^{\text{(r)}}(G_y) \bigg] e^{\text{i} (k_y^{(0)}+G_y) y} \, , \\
\end{aligned}
\end{equation}
where the time dependence is henceforth omitted for the sake of brevity.

The consistent nodal forces produced by the incident, reflected, and transmitted pressure waves, can then be computed as
\begin{equation} \label{f_sl_sr}
    F = \sum_{\text{S}_\text{l}} \int_{\Omega} \, \mathbf{N}_x^T P_{\text{in}} \, d\Omega + 
        \sum_{\text{S}_\text{r}} \int_{\Omega} \, \mathbf{N}_x^T P_{\text{out}} \, d\Omega \, ,
\end{equation}
where $\sum_{\text{S}_\text{l}}$ and $\sum_{\text{S}_\text{r}}$ refer to the FE assembly process performed over the left and right edges of the periodic domain, and $\mathbf{N}_x$ is a matrix containing the $x$-direction interpolation shape functions for the plane elements. Combining Eqs.~(\ref{pin_pout}) and (\ref{f_sl_sr}) and recalling that $\hat{P}_i(G_y)=0$ for $G_y \neq 0$ (case of a single incident plane wave) leads to
\begin{equation} \label{f_fourier1}
\begin{aligned}
    F = \bigg( \sum_{\text{S}_\text{l}} \int_{\Omega} 2 \mathbf{N}_x^T e^{\text{i} k_y^{(0)} y}
    \, d\Omega \bigg) \hat{P}_i(G_y=0)
      &+ \sum_{G_y} \sum_{\text{S}_\text{l}} \text{i} \frac{\rho_e \omega^2}{k_{xa}(G_y)}
      \int_{\Omega} \mathbf{N}_x^T \hat{u}_x^{\text{(l)}}(G_y) e^{\text{i} (k_y^{(0)}+G_y) y} d\Omega \\
      &+ \sum_{G_y} \sum_{\text{S}_\text{r}} \text{i} \frac{\rho_e \omega^2}{k_{xa}(G_y)}
      \int_{\Omega} \mathbf{N}_x^T \hat{u}_x^{\text{(r)}}(G_y) e^{\text{i} (k_y^{(0)}+G_y) y} d\Omega \, .
\end{aligned}
\end{equation}

In the previous equation, the first term corresponds to the immediate force exerted by the incident acoustic wave, while the second and third terms account for the additional fluid loading due to the reflected and transmitted acoustic waves, respectively. Particular attention is given to these terms due to integrals performed over the input and ouptut (left and right, respectively) domains.


The spatial Fourier components of the $x$-direction displacements ($\hat{u}_x(G_y)$) can be computed by applying the orthogonality property in the last equation presented in Eq.~(\ref{pu_fourier}) to obtain
\begin{equation}
    \hat{u}_x(G_y) = \frac{1}{a} \int_{\Omega} u_x \, e^{-\text{i} (k_y^{(0)}+G_y) y} \, d\Omega \, ,
\end{equation}
where $\Omega$ is a edge of the unit cell (e.g., left, right). This equation can be rearranged considering a FE-interpolation in the form
\begin{equation}
    u_x = \mathbf{N}_x \mathbf{u} \, ,    
\end{equation}
where $\mathbf{N}_x$ is a matrix containing the $x$-direction interpolation shape functions and $\mathbf{u}$ is a vector containing the nodal displacements corresponding to the spatial Fourier coefficients. Thus, the previous equation can be rewritten as
\begin{equation}
    \hat{u}_x(G_y) = \frac{1}{a} \mathbf{B}(G_y) \mathbf{u} \, ,
\end{equation}
where $\mathbf{B}(G_y)$ is a matrix relating nodal displacements and spatial Fourier components, given by
\begin{equation}
    \mathbf{B}(G_y) = \sum_{\text{S}} \int_{\Omega} \mathbf{N}_x e^{-\text{i} (k_y^{(0)}+G_y) y} \, d\Omega \, ,
\end{equation}
with $\sum_{\text{S}}$ representing the assembly process considering the edge of the elements at a given face S.
This expression can be substituted into Eq.~(\ref{f_fourier1}), which allows it to be rewritten as
\begin{equation}
    F = F_p + \omega^2 ( \mathbf{M}_f^{\text{in}} + \mathbf{M}_f^{\text{out}} ) \mathbf{u} \, ,
\end{equation}
where $F_p = \big( \sum_{\text{S}_\text{l}} \int_{\Omega} 2 \mathbf{N}_x^T e^{\text{i} k_y^{(0)} y } \, d\Omega \big) \hat{P}_i(G_y=0)$ and the fluid-loading associated matrices written as
$\mathbf{M}_f^{\text{in}} = \frac{\text{i} \rho_e}{a} \sum_{G_y} \frac{1}{k_{xa}(G_y)} \mathbf{B}_{\text{in}}^H(G_y) \mathbf{B}_{\text{in}}(G_y) $ and
$\mathbf{M}_f^{\text{out}} = \frac{\text{i} \rho_e}{a} \sum_{G_y} \frac{1}{k_{xa}(G_y)} \mathbf{B}_{\text{in}}^H(G_y) \mathbf{B}_{\text{out}}(G_y) $, with the matrices $\mathbf{B}_{\text{in}}(G_y)$ and $\mathbf{B}_{\text{out}}(G_y)$ evaluated, respectively, at the incident (left, $\text{S}=\text{S}_{\text{l}}$) and transmitted (right, $\text{S}=\text{S}_{\text{r}}$) faces.


The next step to obtain the STL curves is now considering a periodic medium in the $y$-direction. Notice that the structure may contain fluid-filled acoustic cavities (blue regions in Fig.~\ref{figure_s1}), in which case, its dynamic stiffness matrices can be written partitioned as~\cite{mencik2007wave}
\begin{equation}
    \left[
        \begin{array}{cc}
            \mathbf{K}_s - \omega^2 \mathbf{M}_s & \text{i} \omega \rho_i \mathbf{C} \\
            \text{i} \omega \rho_i \mathbf{C}^T & -\rho_i ( \mathbf{K}_a - \omega^2 \mathbf{M}_a )
        \end{array}
    \right]
    \left\{
    \begin{array}{c}
         \mathbf{u}  \\
         \bm{\psi}
    \end{array}
    \right\}
    =
    \left\{
    \begin{array}{c}
         \mathbf{F}_s  \\
         \frac{1}{\text{i}\omega} \mathbf{F}_a 
    \end{array}
    \right\}
\end{equation}

\section{Comparison between results}
\label{sec_additional_results}

\begin{figure}[h]
 \centering
 \includegraphics[scale=1]{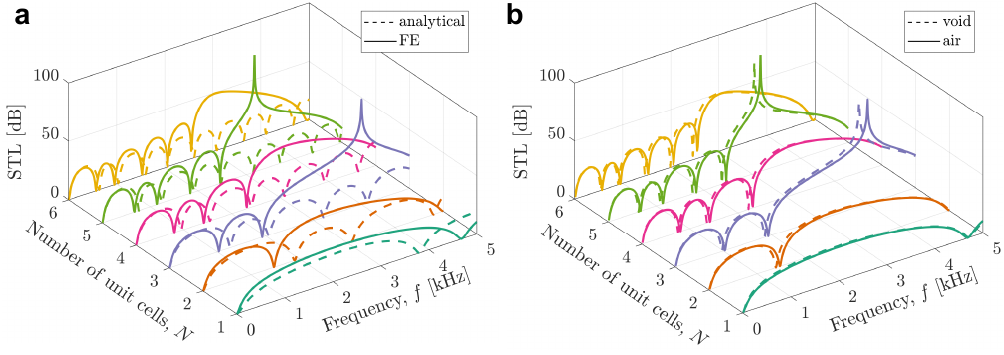}
 \caption{
 Comparison between STL curves obtained using
 (\textbf{a}) analytical and FE-based methods and 
 (\textbf{b}) FE-based methods without (void) and with air in the internal cavities of the metabarrier unit cell.
 }
 \label{figure_s2}
\end{figure}

Figure~\ref{figure_s2} compares the STL curves computed considering different methods and the existence of air-filled cavities in the metabarrier unit cell.
Fig.~\ref{figure_s2}\textbf{a} shows the STL curves computed using the analytical method for homogenized structures (Section~\ref{sec_tmm}) and numerical FE-based method for architected structures (\ref{sec_stl_method_fe}). It is possible to notice that the agreement between both methods is improved as the number of unit cells ($N$) increases, extending the frequency range of agreement between these methods. This observation is in agreement with the hypothesis considered in the homogenization method, which assumes a periodic medium of infinite extent (i.e., the boundary effects, which are more significant for a small number of unit cells, are not considered).
Also, we show in Fig.~\ref{figure_s2}\textbf{b} the differences in the computed STL curves considering vacumm- (void) and air-filled internal cavities of the unit cells of the metabarrier. Notice that in both cases, the external fluid is water. The effect of the inclusion of air in the internal cavities is the redshift of the STL curve, although not significantly.

\section{Optimization results for alternative objective function}
\label{app_fully_coupled}

\begin{figure}[h]
 \centering
 \includegraphics[scale=1]{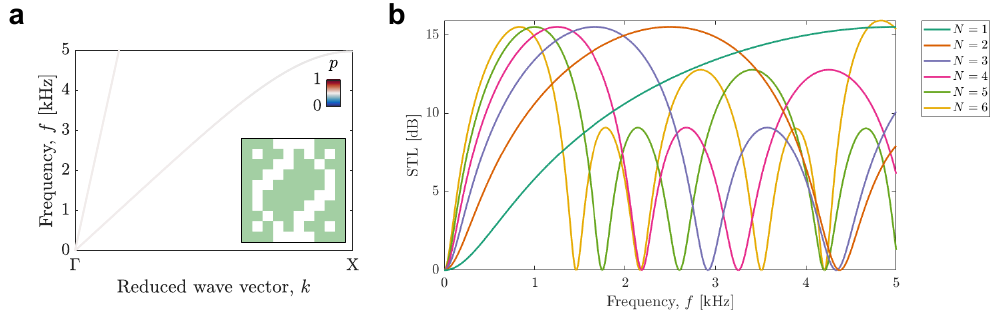}
 \caption{Results for the optimized structure obtained using an alternative objective function.
 \textbf{(a)} Band diagram relative to the obtained structure (inset) showing almost ideally hybridized longitudinal and shear polarization ($p \approx 0.5$).
 (\textbf{b}) STL curves obtained for the homogeneous structure with effective properties for an increasing number of unit cells ($N$).}
 \label{figure_s3}
\end{figure}

An alternative optimization approach is proposed to verify that the effect of achieving high STL values is not owed only to the high degree of polarization between longitudinal and transverse modes (i.e., $p \approx 0.5$). In this case, the chosen objective function (see Eq.~(\ref{optimization_problem})) to be minimized was chosen as
$ \varphi(\mathbf{q}) = \max( \cos^2(\phi_1), \cos^2(\phi_2) )$. Thus, the optimization algorithm seeks to bring both $\phi_1$ and $\phi_2$ as close as possible to $0$, i.e., keeping only shear modes. As this would not be a feasible solution, the algorithm yields a structure with $\phi_{1,2} = \pm 3 \pi/4$, i.e., perfectly polarized ($p=0.5$) modes.
The resulting structure obtained through the topological optimization is shown in Fig.~\ref{figure_s3}\textbf{a} (inset) with the corresponding dispersion relation, which confirms the almost ideal degree of polarization ($p \approx 0.5$). 
Conversely, the STL curves considering a strucutre with effective properties (obtained through the homogenization procedure), shown in Fig.~\ref{figure_s3}\textbf{b} for an increasing number of unit cells ($N$), indicate that the peak STL levels are close to $15$ dB.
This value is significantly lower than the ones observed in Fig.~\ref{figure4}\textbf{c}. This difference can be explained due to the lower value of coupling between normal stresses and shear strains (and vice-versa), given in this case by $\delta = 0.9208$.

\clearpage

\end{document}